\documentclass{article}

\usepackage{caption}
\usepackage{subcaption}
  \usepackage[preprint]{neurips_2026}

\usepackage[utf8]{inputenc} 
\usepackage[T1]{fontenc}    
\usepackage{hyperref}       
\usepackage{url}            
\usepackage{booktabs}       
\usepackage{amsfonts}       
\usepackage{nicefrac}       
\usepackage{microtype}      
\usepackage{xcolor}         
\usepackage[normalem]{ulem}

 \usepackage{multirow}
\usepackage{amsmath}
\usepackage{amssymb}
\usepackage{amsthm}      
\usepackage{mathtools}   
\usepackage{bm}          
\usepackage{bbm}         
\usepackage{centernot}   
\usepackage[ruled,vlined]{algorithm2e} 
\usepackage{enumerate}
\usepackage{enumitem}
\usepackage{wrapfig}

\usepackage{chngcntr}

\counterwithin{table}{section}
\counterwithin{figure}{section}

\newcommand{\bbP}{\mathbb{P}} 
\newcommand{\bbS}{\mathbb{S}} 

\def\ind{\mathbbm{1}} 

\newcommand{\bX}{\mathbf{X}}

\newcommand{\cG}{\mathcal{G}}

\newcommand{\cL}{\mathcal{L}}

\def\Pa{\mathrm{Pa}} 

\def\sX{\mathsf{X}}

\SetKwInput{KwInput}{Input}
\SetKwInput{KwOutput}{Output}

\theoremstyle{remark}
\newtheorem{remark}{Remark}
\newtheorem{assumption}{Assumption}

\newtheorem{theorem}{Theorem}
\newtheorem{corollary}{Corollary}

\hypersetup{hidelinks}

\title{ZIO: Order-Based Structure Learning for Zero-Inflated Count Data under Data Heterogeneity}

\author{%
  Fariha Taskin\\
  Department of Mathematical Sciences\\
  The University of Texas at Dallas\\
  Richardson, TX 75080-3021 \\
  \texttt{Fariha.Taskin@UTDallas.edu} \\
  \And
   Junsouk Choi \\
   Department of Statistics \\
   Korea University\\
   Seoul, 02841, Republic of Korea \\
   \texttt{junsoukchoi@korea.ac.kr} \\
   \AND
   Hyunwoong Chang\thanks{Corresponding author.} \\
   Department of Mathematical Sciences\\
   The University of Texas at Dallas\\
   Richardson, TX 75080-3021 \\
  \texttt{hwchang@utdallas.edu} \\
}

\begin{document}

\maketitle

\begin{abstract}

We propose a structure learning method for directed acyclic graph (DAG) models of zero-inflated multivariate count data. Although causal structure learning with zero-inflated Poisson models is identifiable, existing algorithms rely on local search over the DAG space, leading to poor scalability and a tendency to converge to suboptimal solutions. To address these limitations, we develop a novel order-based scoring framework that searches over variable orderings rather than directly over DAGs. Building on this framework, we further extend the method to multiple heterogeneous datasets by identifying a shared variable ordering while allowing each dataset to have its own edge set. We implement a stochastic hill climbing algorithm with a random-to-random (R2R) proposal operator to efficiently explore the ordering space. Simulation studies show competitive performance against existing methods and improved graph estimation accuracy as more heterogeneous datasets are jointly analyzed under a shared causal ordering. We apply the method to single-nucleus RNA sequencing data from a major depressive disorder study.

\end{abstract}

\section{Introduction}

We consider causal structure learning of a directed acyclic graph (DAG) model for zero-inflated multivariate count data. Such data are common across economics \citep{cameron2013regression}, social sciences \citep{hilbe2011negative}, and biology \citep{yang2022modeling}. One prominent example is single-cell RNA sequencing data, where raw counts are usually log-transformed and modeled as if they follow a Gaussian distribution. However, this practice can introduce bias and lead to inaccurate estimates of causal structure~\citep{gallopin2013hierarchical,nguyen2023structure}. Furthermore, structure learning under Gaussian assumptions is not distributionally identifiable, as multiple distinct graphs can fit the data equally well, posing a major challenge for causal interpretation. 

A structure learning method based on zero-inflated Poisson models has been developed to directly handle zero-inflated count data~\citep{choi2020bayesian}, and the true DAG has been shown to be identifiable. However, the algorithm exhibits suboptimal performance in identifying the true DAG. The approach follows a standard local sampling scheme that explores the graph space via edge additions, deletions, and reversals~\citep{madigan1995bayesian}. Although parallel tempering~\citep{geyer1991markov} is employed to improve convergence~\citep{choi2020bayesian}, the method still suffers from slow convergence and limited scalability. Searching over the space of DAGs is inherently difficult, since the number of possible graphs grows super-exponentially with the number of nodes, for example, the number of labeled DAGs on 10 nodes already exceeds $10^{15}$. While subsequent model extensions~\citep{choi2023model} enhance distributional flexibility, no improved computational methods have been proposed, and these approaches still rely on greedy search over the DAG space.

To address these challenges, we propose an order-based structure learning method for zero-inflated Poisson models. Rather than searching directly over DAGs, order-based methods operate on the permutation space of variables, which is known to be effective due to its smaller search space and the smoother likelihood landscape~\citep{friedman2003being}. Another advantage of the order-based approach is its ability to accommodate datasets from multiple sources. For example, in single-cell RNA sequencing studies, datasets collected from different individuals or experimental conditions often exhibit substantial heterogeneity. Ignoring this heterogeneity and treating all observations as independent and identically distributed from a single latent DAG may lead to misleading results, including inflated false positive edge discoveries. The proposed zero-inflated order-based learning, termed \textbf{ZIO}, accommodates such heterogeneity by allowing source-specific causal structures while assuming a shared topological ordering. The common ordering assumption is widely used in the multiple-DAG estimation literature \citep{wang2020high, li2024bayesian, chang2026leveraging} and is often appropriate in practice. Under this assumption, datasets may have different graph structures, but the causal directions among variables are assumed to remain unchanged. This setting can be reasonable when interventions or environmental shifts are expected to affect the existence or strength of causal links without reversing their direction, as in time-ordered or hierarchical systems. Within this framework, increasing the number of jointly analyzed datasets  can improve DAG estimation accuracy, since joint estimation leverages common ordering signals across datasets and yields more stable graph estimates.

When searching over discrete spaces, the choice of local moves plays an important role in algorithmic efficiency~\citep{zhou2023complexity,chang2026dimension}. To explore the space of orderings, we implement a stochastic hill climbing algorithm using local moves induced by the random-to-random (R2R) operator~\citep{diaconis1993comparison}. Recently, the R2R operator has been adopted in order-based structure learning, where it has been shown to facilitate efficient exploration of the ordering space~\citep{chang2025identifiability, chang2026leveraging}. 
Through extensive experiments, our order-based structure learning method with the R2R operator consistently matches or outperforms existing methods, with particularly strong performance in settings with a larger number of nodes (Section~\ref{Section:Simulation 1}). We also compare the R2R operator with two alternative operators (Section~\ref{subsec:operators}). We further examine how performance varies with different zero proportions in Section~\ref{subsec:zero_rate}.

Finally, we apply our method to single-nucleus RNA sequencing data from a major depressive disorder study~\citep{nagy2020single}. This framework is particularly well suited to case-control single-cell RNA sequencing studies, where the data are inherently zero-inflated and naturally organized across multiple levels, such as patients and cell types. By directly modeling zero-inflated counts without relying on log-Gaussian transformations, our approach enables the characterization of differences in gene regulatory systems between diseased and healthy individuals. We summarize the estimated networks using group-specific edge-selection frequencies to identify shared and group-enriched connections. Weighted total connectivity further summarizes these patterns at the gene level. We also construct consensus networks by retaining edges repeatedly selected within each group and compare their degree rankings. Together, these analyses characterize shared network structure and descriptive differences between controls and cases, providing connections and genes for further biological investigation.

\section{Preliminaries}

\subsection{Structure learning of DAG models}

Structure learning for DAG models is a model selection problem that seeks a DAG representing the dependence relationships among a collection of random variables~\citep{pearl2009causality, spirtes2000causation}. Edge directions often admit a causal interpretation, especially when the data generating process is modeled using recursive structural equations, where each variable is expressed as a function of its parent variables and an exogenous error term. In general, observational data identify the DAG only up to a Markov equivalence class \citep{Koller2009-xc}, and identifiability requires additional assumptions~\citep{Drton2017-oa}. 

\textbf{Structure learning on count data.} DAG learning for count data has been developed through several model classes.
Poisson DAG models and overdispersion scoring provide identifiable structure learning~\citep{park2015learning}, while generalized hypergeometric DAG models extend identifiable discrete DAG learning to a broader family of count distributions~\citep{park2019identifiability}. For zero-inflated counts, \citet{choi2020bayesian} proposed the zero-inflated Poisson Bayesian network (ZIPBN), proving identifiability from observational data. This line of work has been extended to zero-inflated generalized hypergeometric DAG models~\citep{choi2023model}. Recent work has also studied differential zero-inflated count DAGs for two-sample single-cell applications~\citep{choi2025bayesian} and continuous optimization approaches for zero-inflated count DAG learning~\citep{sato2025dag}.

\textbf{Order-based structure learning.} A major computational challenge in DAG learning is that the space of $p$-vertex DAGs grows super-exponentially with the number of nodes $p$. Instead of directly searching over the space of DAGs, order-based methods search the space of variable orderings, that is, the permutation space $\bbS_p$. Given a candidate ordering, the acyclicity constraint is automatically satisfied by restricting the candidate parents of each node to variables that precede it in the ordering. This reformulates graph learning as a collection of local parent-selection problems. 
Order-based learning has been explored in both Bayesian and frequentist contexts~\citep{friedman2003being, teyssier2005ordering,  silander2012simple, agrawal2018minimal, Chang2022-id, li2024bayesian, chang2025identifiability}. To the best of our knowledge, order-based learning has not yet been studied for zero-inflated count data.

\textbf{Multiple DAG estimation.}
The literature on multiple DAG estimation considers the joint learning of  DAGs across heterogeneous datasets by encouraging similarity among graphs while still allowing structural differences~\citep{oates2016exact, lee2022bayesian, castelletti2020bayesian}. A frequently used assumption for inducing such similarity is the common ordering assumption, which assumes that all datasets share the same causal ordering~\citep{shimizu2012joint, wang2020high, chen2021multi}. 
Differential DAG methods have also been proposed to identify structural changes across groups under the common ordering assumption~\citep{li2024bayesian}. Most of these methods focus on Gaussian or linear structural equation models. To the best of our knowledge, joint order-based learning of multiple zero-inflated count DAGs has not yet been studied.

\subsection{Notation}
Let $G = (V, E)$ denote a DAG, where $V=[p] = \{1, \dots, p\}$ is the vertex set and $E \subset V \times V$ is the set of directed edges, with no directed cycle. Let $|G|$ denote the cardinality of the edge set $E$. If a directed edge is present, we write $(i, j) \in E$ to indicate the edge direction $i \rightarrow j$ in $G$ for $i, j \in V$. For a DAG $G$ and node $j$, we denote the set of parent nodes as $\Pa_j(G)$. Let $\sigma = (\sigma_1,\dots,\sigma_p)$ denote an ordering of the nodes, where $\sigma_j$ denotes the node in the $j$-th position, for $j\in[p]$. It represents a topological ordering in DAG $G$ if and only if, for every pair of indices $i<j$, the node $\sigma_i$ is not a descendant of $\sigma_j$ within $G$. We denote $\bbS_p$ as the set of all orderings in $[p]$. Let $\sigma^{-1}(j)$ denote the position of node $j$ in the ordering $\sigma$, and for each node  $j \in [p]$, 
\begin{align*}
P_j^{\sigma} = \{i: \sigma^{-1}(i) < \sigma^{-1}(j)\}
\end{align*}
denote the set of potential parents of node $j$ under the ordering $\sigma$, i.e., all nodes preceding $j$ in $\sigma$. Let $\cG_p$ be the collection of all $p$-node DAGs and $\cG_p^\sigma$ be the collection of all $p$-node DAGs consistent with topological ordering $\sigma$; i.e., $\cG_p^\sigma = \{G \in \cG_p:\{i \to j\}\in G \text{ implies } \sigma^{-1}(i) < \sigma^{-1}(j)\}$. Let $\mathcal L(G)$ denote the set of all topological orderings of a DAG $G$. 

\section{Order-based structure learning with zero-inflated Poisson models}

\subsection{Model specification}

\textbf{Single DAG estimation.}
We first define the model and the order-based score for a single dataset.
Let $\mathbf X=(x_{ij})\in\mathbb N_0^{n\times p}$ denote the zero-inflated count data matrix with sample size $n$, where each row is an i.i.d. copy of a $p$-dimensional random vector $\sX=(\sX_1,\ldots,\sX_p)$. 
For an ordering $\sigma \in \bbS_p$ and an underlying DAG $G \in \cG_p^\sigma$, we model the conditional probability mass function of $\sX_j$ given $\sX_{\Pa_j(G)}$ using the following zero-inflated Poisson model
\begin{equation}\label{eq:zip}
\begin{aligned}
\bbP\big(\sX_j = x \mid \sX_{\Pa_j(G)}\big) =
\begin{cases}
\eta_j + (1 - \eta_j) e^{-\lambda_j}, & x = 0, \\[6pt]
(1 - \eta_j) e^{-\lambda_j} (\lambda_j)^x/x!, & x > 0,
\end{cases}
\end{aligned}
\end{equation}\\
where \(\text{logit}(\eta_j) = \sum_{h \in \Pa_j(G)} \alpha_{jh} \sX_h + \delta_j\), \(\log(\lambda_j) = \sum_{h \in \Pa_j(G)} \beta_{jh} \sX_h + \gamma_j\), and $\sX_{\Pa_j(G)} = \{\sX_h:h\in \Pa_j(G)\}$.
Here, the parameter $\eta_j$ indicates the extra probability mass at zero beyond the zeros from the Poisson component, and the parameter $\lambda_j$ represents the Poisson rate parameter. 
From~\eqref{eq:zip}, it is clear that $h \notin \Pa_j(G)$ if and only if $\alpha_{jh} = \beta_{jh} = 0$, so an edge $h \to j$ exists if and only if $(\alpha_{jh}, \beta_{jh}) \neq (0, 0)$. We define the parameter vector for node $j$ under parent set $S$ as
\[
\theta_j(S)
=
\left(
\{\alpha_{jh}:h\in S\},
\{\beta_{jh}:h\in S\},
\delta_j,\gamma_j
\right),
\]
and let $\ell_j(S,\theta_j(S);\mathbf X_j,\mathbf X_S)$ denote the node-wise log-likelihood function. For fixed $S$, the maximum likelihood estimator of $\theta_j(S)$ is given by $\widehat \theta_j(S)
\in
\arg\max_{\theta_j(S)}
\ell_j(S,\theta_j(S);\mathbf X_j,\mathbf X_S)$. For a given ordering $\sigma$, 
we select the parent set of node $j$ by 
\begin{align}\label{eq:var_sel}
    \widehat S_j(\sigma)
\in
\arg\max_{S\subseteq P_j^\sigma}
\left[
2\ell_j(S,\widehat\theta_j(S);\mathbf X_j,\mathbf X_S)
-
c_0 d_j(S)\log n
\right],
\end{align}
where $c_0>0$ is a penalty constant and $d_j(S)=\dim\{\theta_j(S)\}=2|S|+2$. We define the local score for node $j$ given $\sigma$ as
\[
\mathrm{OBS}_j(\sigma)
=
2\ell_j(\widehat S_j(\sigma),\widehat\theta_j(\widehat S_j(\sigma));
\mathbf X_j,\mathbf X_{\widehat S_j(\sigma)})
-
c_0 d_j(\widehat S_j(\sigma))\log n,
\]
and the order-based score for an ordering $\sigma$ is $\mathrm{OBS}(\sigma)
=
\sum_{j=1}^p \mathrm{OBS}_j(\sigma).$ The estimated DAG under ordering $\sigma$ is denoted by $\widehat G_\sigma$, with edge set $\widehat E_\sigma
=
\{h\to j:h\in \widehat S_j(\sigma),\ j\in[p]\}.$

Next, we show that the true causal ordering can be identified under the proposed order-based model. Suppose that the dataset $\mathbf X$ is generated by~\eqref{eq:zip} associated with the true DAG $G^*$ with the true parameter $\Theta^* = \{ \theta_j^*\}_{j=1}^p$, where $\theta_j^* =
(
\{\alpha_{jh}^*:h\in \Pa_j(G^*)\},
\{\beta_{jh}^*:h\in \Pa_j(G^*)\},
\delta_j^*,\gamma_j^*
)$ for $j \in [p]$.
We introduce the following assumptions. 
\begin{assumption}\label{assump:p_fixed}
The number of nodes \(p\) is fixed, and the maximum in-degree of each node in \(G^{*}\) is uniformly bounded and does not grow with the sample size \(n\).
\end{assumption}

\begin{assumption}[Likelihood regularity]
\label{assump:lik_regular}
For every \(G\in\mathcal G_p\), the parameter space \(\Omega_G\) is such that
the MLE exists with probability tending to one and the expected log-likelihood
\(
\widetilde\ell_G(\Theta_G)
=
\mathbb{E}_{p^*}\log p_G(X;\Theta_G)
\)
has a finite, attained maximum. 
The log-likelihood class
\(
\{\log p_G(\cdot;\Theta_G):\Theta_G\in\Omega_G\}
\)
is \(p^*\)-Glivenko--Cantelli, so that
$
\sup_{\Theta_G\in\Omega_G}
\big|
n^{-1}\ell_{G,n}(\Theta_G)
-
\widetilde\ell_G(\Theta_G)
\big|
\xrightarrow{p}0
$. \\
Moreover, for every overfitted graph \(G \supsetneq G^*\), the model under \(G^*\) is a regular nested submodel of the model under \(G\), and the standard regularity conditions for Wilks' theorem hold for the restrictions corresponding to the extra edges in \(E(G)\setminus E(G^*)\).
\end{assumption}

\begin{assumption}[Causal minimality]
\label{assump:minimal}
The data are generated from the ZIPBN model $p^*(x)=p_{G^*}(x;\Theta^*)$, where the true DAG \(G^*\) is causally minimal. That is, for every true edge $h \to j \in E(G^*)$, $(\alpha_{jh}^*, \beta_{jh}^*) \ne (0, 0)$.
\end{assumption}

\begin{theorem}\label{thm:consistency}
Suppose that Assumptions~\ref{assump:p_fixed}--\ref{assump:minimal} hold for the dataset $\bX$, the true data generating DAG \(G^{*}\) and the true parameter $\Theta^*$. For every consistent ordering  \(\sigma \in \mathcal{L}(G^{*}),\) (1) $\bbP\{\mathrm{OBS}(\sigma)>\mathrm{OBS}(\tau)\}\to 1$ for any ordering \(\tau \notin \mathcal{L}(G^{*}),\) (2) $\bbP\{\widehat G_\sigma=G^*\}\to 1$, as $n \to \infty$. 
\end{theorem}

All proofs are deferred to Section~\ref{sec:proofs} of the supplementary material. We also provide model-specific likelihood regularity conditions for Assumption~\ref{assump:lik_regular} (Section~\ref{subsec:alt_assump2}). 
We note that the result assumes optimal parent selection for each ordering, whereas our implementation uses forward–backward stepwise selection as an approximation. This gap is examined in Section~\ref{subsec:optimization} of the supplementary material, which empirically shows that the stepwise approximation generally agrees with optimal parent selection.

\textbf{Multiple DAG estimation.} We extend the framework to $K$ datasets. For $k \in [K]$, let $\bX^{(k)} = (x^{(k)}_{ij})\in \mathbb{N}_0^{n_k \times p}$ denote the $k$-th zero-inflated count data matrix with sample size $n_k$. We note that all datasets contain the same $p$ variables, and each row of $\bX^{(k)}$ is an i.i.d. copy of $\sX^{(k)} = ( \sX^{(k)}_1, \sX^{(k)}_2, \ldots,\sX^{(k)}_p)$, which follows the probability distribution described in \eqref{eq:zip} associated with the underlying DAG $G^{(k)}$ and the parameter $\Theta^{(k)} = \{\theta_j^{(k)}\}_{j=1}^p$. We assume that, for each $k\in [K]$, Assumptions~\ref{assump:p_fixed}--\ref{assump:minimal} hold for the dataset $\bX^{(k)}$ and ($G^{(k)}$, $\Theta^{(k)}$). As previously defined, let $\mathrm{OBS}^{(k)}(\sigma)$ denote the order-based score for the $k$-th dataset under ordering $\sigma$, and define the joint order-based score as $\mathrm{JOBS}(\sigma) = \sum_{k=1}^K \mathrm{OBS}^{(k)}(\sigma)$. We further assume the common ordering assumption.
\begin{assumption}\label{assump:a3}
    There exists a permutation $\sigma \in \bbS_p$ that satisfies $\sigma \in \cL(G^{(k)})$ for all $k \in [K]$, that is, $\bigcap_{k=1}^{K}\cL(G^{(k)}) \neq \emptyset$.
\end{assumption}

\begin{corollary}\label{cor:multi}
Suppose that Assumptions~\ref{assump:p_fixed}--\ref{assump:minimal} hold for each dataset $\bX^{(k)}$, with true data generating DAG $G^{(k)}$ and true parameter $\Theta^{(k)}$, for all $k\in[K]$. Suppose that Assumption~\ref{assump:a3} holds, and $K$ is a fixed constant. For every common ordering $\sigma \in \bigcap_{k=1}^{K}\cL(G^{(k)})$, (1) $\bbP\{\mathrm{JOBS}(\sigma) > \mathrm{JOBS}(\tau)\} \to 1$ for $\tau \notin \bigcap_{k=1}^{K}\cL(G^{(k)})$, and (2) $\bbP\{\bigcap_{k=1}^K \{\hat{G}^{(k)}_\sigma = G^{(k)}\}\} \to 1$ as $\min_{k\in[K]} n_k \to \infty$.
\end{corollary}

\begin{remark}\label{rmk:multiple}
    We observe improved estimation accuracy for the true graphs under joint estimation  (Section~\ref{subsec:performance}). This is because the joint
score aggregates directional information across the $K$ datasets. If an ordering is inconsistent with the true graph of at least one dataset, that dataset contributes a score gap relative to a common ordering. By summing the source-specific scores, these gaps accumulate across datasets, making inconsistent orderings easier to distinguish from consistent orderings. Thus, additional heterogeneous datasets can improve order recovery. We note that the computation is naturally parallelizable with multiple cores, since individual order-based scores can be evaluated independently.
\end{remark}

\subsection{Hill climbing with the random-to-random operator}
We consider a stochastic hill climbing algorithm to find solutions in the permutation space $\bbS_p$. 
At each iteration, the algorithm generates a candidate ordering by applying a local move operator to the current ordering. While there are many local move operators, we use the random-to-random operator, which is defined as
\begin{align*}
\operatorname{R2R}(\sigma, i, j) =
    \begin{cases}
    (\sigma_1, \dots,\sigma_{j-1},\sigma_i, \sigma_j,\dots, \sigma_{i-1}, \sigma_{i+1}, \dots, \sigma_p), \ j < i,\\
    (\sigma_1, \dots,\sigma_{i-1},\sigma_{i+1}, \dots, \sigma_{j}, \sigma_{i},\sigma_{j+1}, \dots, \sigma_p), \ i < j,
\end{cases}
\end{align*}
for two distinct positions $i,j\in[p]$. Our choice of R2R is motivated by two complementary Gaussian results: \citet{chang2025identifiability} establish the absence of strict local optima under a minimum-trace Gaussian SEM, while \citet{chang2026leveraging} show that
R2R moves admit non-decreasing individual score paths through covered edge reversals followed by edge deletions~\citep{chickering2002optimal}. We note that these guarantees do not directly extend to our setting, and a corresponding result remains open.
\begin{algorithm}[!b]
\captionsetup{font={normalsize,stretch=1}}
\caption{Stochastic hill climbing with random-to-random (R2R) proposal operator}
\label{alg1}
\DontPrintSemicolon
\LinesNotNumbered     
\SetKwInOut{KwIn}{Input}
\SetKwInOut{KwOut}{Output}

\KwIn{ Datasets $\{\mathbf{X}^{(k)}\}_{k=1}^K$, an initial ordering $\sigma^0$, the number of iterations $T$.}

\(\mathrm{JOBS}_{\sigma^0} \gets \mathrm{JOBS}(\sigma^0)\) \;
\For{$t = 1, \dots, T$}{
     Sample two distinct positions $q,r \in [p]$ uniformly at random\;
    $\sigma' \gets \mathrm{R2R}(\sigma^{t-1}, q, r)$  \; 
    \(
      \mathrm{JOBS}_{\sigma'} \gets \mathrm{JOBS}(\sigma')
    \)
    
    \uIf{$\mathrm{JOBS}_{\sigma'} > \mathrm{JOBS}_{\sigma^{t-1}}$}{
        $\sigma^t \gets \sigma'$, \quad $\hat{G}^{(k)}_{\sigma^t} \gets \hat{G}_{\sigma'}^{(k)}$
    }
    \Else{
        $\sigma^t \gets \sigma^{t-1}$ \;
    }
}
\KwOut{ Estimated DAGs $\{\hat{G}_{\sigma^T}^{(k)}\}_{k=1}^K$ and final ordering $\sigma^T$.} 
\end{algorithm}

At each iteration, we sample two distinct positions $i, j \in [p]$ uniformly at random and generate a new candidate order $\sigma' = \operatorname{R2R}(\sigma, i, j)$. Given $\sigma'$, we solve~\eqref{eq:var_sel} to identify the parent set of each node $j$ that maximizes the score. Since exhaustive search over all parent subsets is computationally infeasible for moderate $p$, we use forward--backward stepwise variable selection, in which candidate parents are first added through forward selection and then removed through backward elimination according to the score. For each dataset, this yields the estimated graph $\widehat{G}_{\sigma'}$ associated with the candidate ordering $\sigma'$, together with its corresponding score $\mathrm{OBS}(\sigma')$. The joint score $\mathrm{JOBS}(\sigma')$ across all datasets is then computed by combining the order-based scores across datasets. If no stopping rule is used, this process is repeated for a fixed number of iterations, resulting in an estimated variable ordering and the corresponding graph. We present an overview of the proposed procedure in Algorithm~\ref{alg1}. The R code for the proposed method and simulation studies can be found at \url{https://github.com/Fariha-taskin/zio}.

\section{Simulation study}
\label{Section:Simulation} 

\subsection{Performance evaluation}\label{subsec:performance}
\textbf{On single dataset:}
\label{Section:Simulation 1} 
We evaluate the performance of the proposed method using synthetic data and compare it with other methods. We follow the simulation setup of \cite{choi2020bayesian}. In the simulation, we consider the sample size $n = 1000$, and different numbers of nodes $p \in \{25, 50, 75\}$. 
We generate each DAG by sampling exactly $p$ directed edges without introducing directed cycles.
To obtain simulated datasets with approximately $50\%$ zeros, we sample the edge-specific parameters and intercepts from the distributions specified in Table~\ref{tab:zero_parameter_ranges} of the supplementary material. Unless otherwise stated, simulation experiments use 30 replicates.
\begin{table}[!b]
\vspace{-10pt}
\centering
\captionsetup{font={normalsize,stretch=1}}
\caption{Comparison of methods using TPR and FDR for $K=1$, $n=1000$,
$p\in\{25,50,75\}$, and $50\%$ zeros, following~\cite{choi2020bayesian}.
Each entry reports the mean (standard error) based on 30 experiments.
The best performance is in boldface.}
\label{tab:comparison_vertical}
\begingroup
\renewcommand{\arraystretch}{1.3}
\resizebox{\linewidth}{!}{%
\begin{tabular}{cccccccc} 
\toprule
 $p$ & Metric & ZIO & ZIPBN & ODS & MRS & ZICO & NOTEARS\\
\midrule
\multirow{2}{*}{25} 
& TPR & 0.917 (0.016) & 0.821 (0.015) & 0.370 (0.018) & 0.812 (0.013) & \textbf{0.938 (0.010)} & 0.900 (0.016) \\
& FDR & \textbf{0.104 (0.014)} & 0.216 (0.019) & 0.695 (0.013) & 0.324 (0.015) & 0.130 (0.016) &  0.114 (0.018) \\
\midrule
\multirow{2}{*}{50} 
& TPR & \textbf{0.905 (0.010)} & 0.781 (0.016) & 0.386 (0.014) & 0.804 (0.010) & 0.887 (0.010) & 0.864 (0.013) \\
& FDR & \textbf{0.122 (0.009)} & 0.255 (0.017) & 0.689 (0.011) & 0.435 (0.001) & 0.223 (0.017) & 0.131 (0.014) \\
\midrule
\multirow{2}{*}{75} 
& TPR & \textbf{0.903 (0.009)} & 0.758 (0.013) & 0.344 (0.004) & 0.727 (0.004) & 0.838 (0.008) & 0.820 (0.009) \\
& FDR & \textbf{0.139 (0.007)} & 0.281 (0.013)  & 0.733 (0.007) & 0.561 (0.006) & 0.315 (0.012) & 0.195 (0.012) \\
\bottomrule
\end{tabular}%
}
\endgroup
\end{table}
For the proposed method, we set $c_0=1$ as the hyperparameter throughout the paper; this choice corresponds to the negative Bayesian information criterion. We use $T \in \{1000, 3000, 3500\}$ iterations for $p \in \{25, 50, 75\}$, respectively.
We evaluate five baseline algorithms: ZIPBN~\citep{choi2020bayesian}, ODS~\citep{park2015learning}, MRS~\citep{park2019identifiability}, ZICO~\citep{sato2025dag}, and NOTEARS~ \citep{zheng2018dags}. The hyperparameter settings used for the competing methods are reported in Table~\ref{tab:method_hyperparameters} of the supplementary material.  We summarize performance using standard metrics, including the true positive rate (TPR) and false discovery rate (FDR), providing a clear assessment of how the proposed method compares with existing approaches. The results for each scenario are reported in Table~\ref{tab:comparison_vertical}. 
Overall, the proposed method remains competitive at $p=25$, and achieves the highest TPR and lowest FDR at the larger $p$ settings.

\textbf{On multiple datasets:}
\label{Section:Simulation 2}
\begin{table}[!t]
\captionsetup{font={normalsize,stretch=1}}
\vspace{-5mm}
\caption{Average edge difference between the true and estimated DAGs, average Kendall's tau between the true and estimated orderings, TPR, and FDR for varying $K$ and $p$, with $n = 1000$, and $ 50\%$ zeros. Each entry reports the mean (standard error) based on 30 experiments.}
\label{tab:performance}
\centering
\small
\renewcommand{\arraystretch}{1.3}
\begin{tabular}{cccccc}
\toprule
\multirow{2}{*}{} & \multicolumn{5}{c}{$p$} \\
\cmidrule(lr){3-6}
 $K$ & Metric & 5 & 10 & 20 & 40\\ 
\midrule 
\multirow{4}{*}{1} 
& $\Delta$ & 0.267 (0.185) & 1.200 (0.334) & 2.633 (0.506) & 7.967 (0.841) \\
 & $\tau$ & 0.647 (0.055) & 0.406 (0.039) & 0.321 (0.023) & 0.311 (0.016) \\
 & TPR & 0.973 (0.019) & 0.923 (0.023) & 0.935 (0.014) & 0.904 (0.012) \\
 & FDR & 0.027 (0.019) & 0.044 (0.015) & 0.064 (0.014) & 0.100 (0.012) \\
\midrule
\multirow{4}{*}{5} 
& $\Delta$ & 0.034 (0.028) & 0.513 (0.077) & 1.773 (0.328) & 5.948 (0.385) \\
 & $\tau$ & 0.986 (0.010) & 0.923 (0.014) & 0.814 (0.018) & 0.740 (0.013) \\
 & TPR & 0.997 (0.003) & 0.959 (0.007) & 0.948 (0.012) & 0.924 (0.005) \\
 & FDR & 0.004 (0.003) & 0.011 (0.003) & 0.037 (0.006) & 0.072 (0.006) \\
\midrule
\multirow{4}{*}{10} 
&$\Delta$ & 0.047 (0.021) & 0.443 (0.046) & 1.410 (0.108) & 4.191 (0.143) \\
 & $\tau$ & 0.993 (0.007) & 0.986 (0.005) & 0.936 (0.007) & 0.879 (0.005) \\
 & TPR & 0.994 (0.003) & 0.965 (0.004) & 0.956 (0.003) & 0.940 (0.003) \\
 & FDR & 0.003 (0.002) & 0.009 (0.002) & 0.026 (0.003) & 0.045 (0.002) \\
\midrule
\multirow{4}{*}{20} 
& $\Delta$ & 0.052 (0.013) & 0.418 (0.038) & 1.363 (0.077) & 4.062 (0.184) \\
 & $\tau$ & 1.000 (0.000) & 1.000 (0.000) & 0.984 (0.003) & 0.953 (0.003) \\
 & TPR & 0.992 (0.002) & 0.968 (0.003) & 0.954 (0.002) & 0.942 (0.004) \\
 & FDR & 0.002 (0.001) & 0.010 (0.001) & 0.022 (0.002) & 0.044 (0.002) \\
\bottomrule
\end{tabular}
\end{table}
To evaluate the performance of the proposed method in a heterogeneous setting, we simulate data across $K \in \{1,5,10,20\}$ datasets and vary the number of nodes as $p \in \{5,10,20,40\}$. For each configuration, we generate $K$ sparse DAGs $\{G^{(k)}\}_{k=1}^K$ sharing the true ordering $\sigma^*=(1,\ldots,p)$ while allowing dataset-specific edge sets. Each DAG contains exactly $p$ distinct directed edges sampled uniformly from those consistent with $\sigma^*$.  
Zero-inflated count data are generated from each DAG using the same parameter ranges as in the single-dataset setting, yielding approximately $50\%$ zeros. We set the number of search iterations to $T=250,500,1500,$ and $3000$ for $p=5,10,20,$ and $40$, respectively. 
To measure estimation accuracy, we use the average difference in edges \((\Delta)\) between the true DAGs   and their estimates  \(\{\hat{G}^{(k)}\}_{k = 1}^K\) 
, defined as \(K^{-1}\sum_{k}\sum_{i,j}|G^{(k)}_{ij} -\hat{G}^{(k)}_{ij}|\). Also, we use Kendall's tau \((\tau)\) between the true and estimated orderings to assess ordering recovery, where a value closer to 1 indicates that the algorithm successfully recovers the causal ordering. In addition to $\Delta$ and $\tau$, we report the true positive rate (TPR) and false discovery rate (FDR). Table~\ref{tab:performance} presents the results. Increasing the number of data sources generally leads to smaller $\Delta$ and larger $\tau$, suggesting that the proposed method improves both graph and ordering estimation accuracy as $K$ increases. 
Finally, wall times are reported in Section~\ref{subsec:walltime} of the supplementary material, and Section~\ref{subsec:shortlist} proposes a strategy to accelerate computation.

\begin{table}[!b]
\vspace{-10pt}
\captionsetup{font={normalsize,stretch=1}}
\caption{Performance of the proposed method using TPR and FDR under different zero proportion levels ($ 25\%$, $ 50\%$, and $ 75\%$), with $p = 20$, $K \in \{1,5\}$ and $n \in \{250,500,1000,2000\}$. Each entry reports the mean (standard error) based on 30 experiments.}

\label{tab:zero_inflation_results}
\centering
\begin{tabular}{ccc cccc}
\toprule
\multirow{2}{*}{} & & & \multicolumn{4}{c}{$n$} \\ 
\cmidrule(lr){4-7}
Zero proportion & $K$ & Metric & 250 & 500 & 1000 & 2000 \\
\midrule

\multirow{4}{*}{$ 25\%$} 
& \multirow{2}{*}{1}  & TPR & 0.903 (0.016) &  0.950 (0.012) &  0.993 (0.003)&  0.993 (0.004)\\
&                     & FDR & 0.188 (0.015) &  0.126 (0.015) &  0.036 (0.007)&  0.037 (0.009)\\
& \multirow{2}{*}{5} & TPR & 0.910 (0.016) &  0.965 (0.011) &  0.992 (0.003)&  0.995 (0.003)\\
&                     & FDR & 0.152 (0.014) &  0.092 (0.012) &  0.029 (0.008)&  0.034 (0.008)\\

\midrule

\multirow{4}{*}{$ 50\%$} 
& \multirow{2}{*}{1}  & TPR & 0.763 (0.027) &  0.860 (0.026)&  0.935 (0.014)&  0.953 (0.012)\\
&                     & FDR & 0.202 (0.020) &  0.130 (0.018)&  0.064 (0.014)&  0.054 (0.011)\\
& \multirow{2}{*}{5} & TPR
& 0.813 (0.024) & 0.925 (0.021) & 0.948 (0.012) & 0.979 (0.007)\\
&                     & FDR & 0.128 (0.015) &  0.061 (0.012)&  0.037  (0.006) & 0.016 (0.007)\\

\midrule

\multirow{4}{*}{$ 75\%$} 
& \multirow{2}{*}{1}  & TPR & 0.597 (0.022) &  0.765 (0.019)&  0.800 (0.024)&  0.937 (0.011)\\
&                     & FDR &  0.411 (0.022)&  0.250 (0.018)&  0.215 (0.023)&  0.077 (0.012)\\
& \multirow{2}{*}{5} & TPR & 0.662 (0.023) &  0.898 (0.013)&  0.930 (0.011)&  0.983 (0.005)\\
&                     & FDR &  0.326 (0.024)&  0.116 (0.014)&  0.082 (0.012)&  0.027 (0.006)\\

\bottomrule
\end{tabular}
\end{table}

\subsection{Sensitivity analysis}\label{subsec:zero_rate}

\textbf{Zero proportion.} We conduct a sensitivity analysis to evaluate the robustness of the proposed method as the proportion of zeros increases, as considered in previous studies~\citep{choi2020bayesian, choi2023model, jun2026zero, choi2017network}. The proportion of zeros can vary substantially across zero-inflated count applications, and higher zero proportions generally make structure learning more difficult because fewer observations contain informative nonzero variation.  We consider datasets with approximately $25\%$, $50\%$, and $75\%$ zeros. We generate datasets with varying proportions of zeros using the parameter ranges specified in Table~\ref{tab:zero_parameter_ranges} of the supplementary material.
We consider $n \in \{250,500,1000,2000\}$, $p=20$, and $K \in \{1,5\}$, with $T=1500$ iterations. As shown in Table~\ref{tab:zero_inflation_results}, graph recovery generally deteriorates as the proportion of zeros increases, while larger sample sizes and joint estimation across multiple datasets improve performance.

\textbf{Likelihood misspecification.}
We also examine sensitivity to likelihood misspecification by fitting ZIP and ZINB models to data generated from both distributions, and report the results in the supplementary material (Section~\ref{subsec:model_misspecification}).


\section{Real data analysis}
\label{sec:real_data}
We consider a major depressive disorder (MDD) case-control dataset generated from publicly available single-nucleus RNA-seq profiling of post-mortem brain samples~\citep{nagy2020single}. The dataset includes 34 individuals: 17 with MDD and 17 healthy controls, with varying sample sizes. One control subject is excluded due to an insufficient number of cells. We retain the 41 genes with an FDR below 0.10, as reported in \citet{nagy2020single}, and use the raw gene count matrices. The sample sizes and the gene list are provided in  Section~\ref{subsec:gene_description} of the supplementary material. 
We run the proposed method for 3000 iterations and obtain a subject-specific DAG estimate $\widehat G^{(k)}$ for each dataset $k=1,\ldots,33$. 
\begin{table}[!b]
\vspace{-12pt}
\centering
\small
\captionsetup{font={normalsize,stretch=1}}
\caption{Top five group-enriched directed edges ranked by the positive difference in group-specific edge-selection frequencies. Selection frequencies are reported with subject counts in parentheses.}
\label{tab:shared_enriched_edges}
\begingroup
\renewcommand{\arraystretch}{1.3}
\begin{tabular}{cllccc}
\toprule
Category & Rank & Directed edge
& $\widehat{\pi}^{(\mathrm{ctrl})}$
& $\widehat{\pi}^{(\mathrm{MDD})}$
& Score \\
\midrule
\multirow{5}{*}{Control-enriched}
& 1 & $\mathrm{CEND1}\to\mathrm{HSP90AA1}$
& 0.875~(14/16) & 0.353~(6/17) & 0.522 \\
& 2 & $\mathrm{PEBP1}\to\mathrm{TAPBP}$
& 0.688~(11/16) & 0.176~(3/17) & 0.511 \\
& 3 & $\mathrm{TSC22D1}\to\mathrm{KAZN}$
& 0.750~(12/16) & 0.294~(5/17) & 0.456 \\
& 4 & $\mathrm{APP}\to\mathrm{ACAA1}$
& 0.625~(10/16) & 0.176~(3/17) & 0.449 \\
& 5 & $\mathrm{APP}\to\mathrm{PINK1}$
& 0.562~(9/16) & 0.176~(3/17) & 0.386 \\
\midrule
\multirow{5}{*}{MDD-enriched}
& 1 & $\mathrm{KAZN}\to\mathrm{KLC2}$
& 0.000~(0/16) & 0.353~(6/17) & 0.353 \\
& 2 & $\mathrm{DOCK4}\to\mathrm{TMSB4X}$
& 0.188~(3/16) & 0.529~(9/17) & 0.342 \\
& 3 & $\mathrm{GRIN2A}\to\mathrm{FAM66C}$
& 0.562~(9/16) & 0.882~(15/17) & 0.320 \\
& 4 & $\mathrm{TSC22D1}\to\mathrm{HIVEP1}$
& 0.062~(1/16) & 0.353~(6/17) & 0.290 \\
& 5 & $\mathrm{TOM1L1}\to\mathrm{KIF26B}$
& 0.062~(1/16) & 0.353~(6/17) & 0.290 \\
\bottomrule
\end{tabular}
\endgroup
\end{table}
To examine group-level patterns in the estimated DAGs, we calculate the selection frequency of each directed edge $i\to j$ within group $g\in\{\mathrm{ctrl}, \mathrm{MDD}\}$, defined as
\[
\widehat{\pi}_{ij}^{(g)}
=
\frac{1}{K_g}
\sum_{k\in g}
\mathbf{1}\!\left\{
i\to j\in\widehat G^{(k)}
\right\},
\]
where $K_{\mathrm{ctrl}}=16$ and $K_{\mathrm{MDD}}=17$. We identify 15 directed edges that are shared in all 33 subjects (see Table~\ref{tab:shared_edges} in the supplementary material).

\noindent
\textbf{Edge-level differences.}
To identify group-enriched connections, we decompose the edge-selection frequencies into shared, control-enriched, and MDD-enriched components, with
\[
\widehat{\pi}_{ij}^{(\mathrm{shared})}
=
\min\!\left\{
\widehat{\pi}_{ij}^{(\mathrm{ctrl})},
\widehat{\pi}_{ij}^{(\mathrm{MDD})}
\right\}, 
\]
\[
\widehat{\pi}_{ij}^{(\mathrm{ctrl\text{-}enriched})}
=
\left(
\widehat{\pi}_{ij}^{(\mathrm{ctrl})}
-
\widehat{\pi}_{ij}^{(\mathrm{MDD})}
\right)_{+},
\qquad
\widehat{\pi}_{ij}^{(\mathrm{MDD\text{-}enriched})}
=
\left(
\widehat{\pi}_{ij}^{(\mathrm{MDD})}
-
\widehat{\pi}_{ij}^{(\mathrm{ctrl})}
\right)_{+},
\]
where $(a)_{+}=\max\{a,0\}$. We rank directed edges within each component and report the largest enrichment scores in Table~\ref{tab:shared_enriched_edges}. The strongest control-enriched edges include $\mathrm{CEND1}\to\mathrm{HSP90AA1}$ and $\mathrm{PEBP1}\to\mathrm{TAPBP}$, where  HSP90AA1 and PEBP1 are known to be involved in cellular signaling~\citep{zuehlke2015regulation,rajkumar2016understanding}. 

\begin{wrapfigure}{r}{0.58\textwidth}
    \centering
    \captionsetup{font={small,stretch=1}}
    \includegraphics[width=\linewidth]{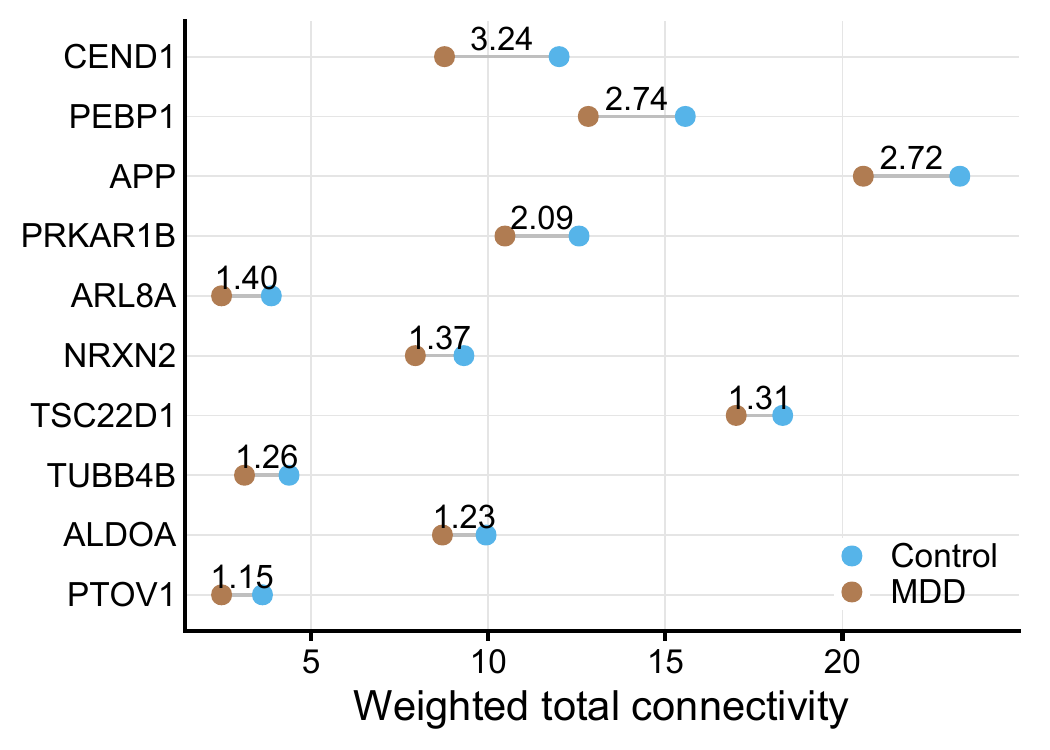}
    \caption{Weighted total connectivity for the ten genes with the largest absolute differences between controls and MDD cases. Numbers show Control--MDD differences.}
    \label{fig:weight_connectivity}
    \vspace{-10pt}
\end{wrapfigure}

\begin{figure}[!b]
    \centering
    \captionsetup{font={normalsize,stretch=1}}
    \includegraphics[width=0.49\linewidth]{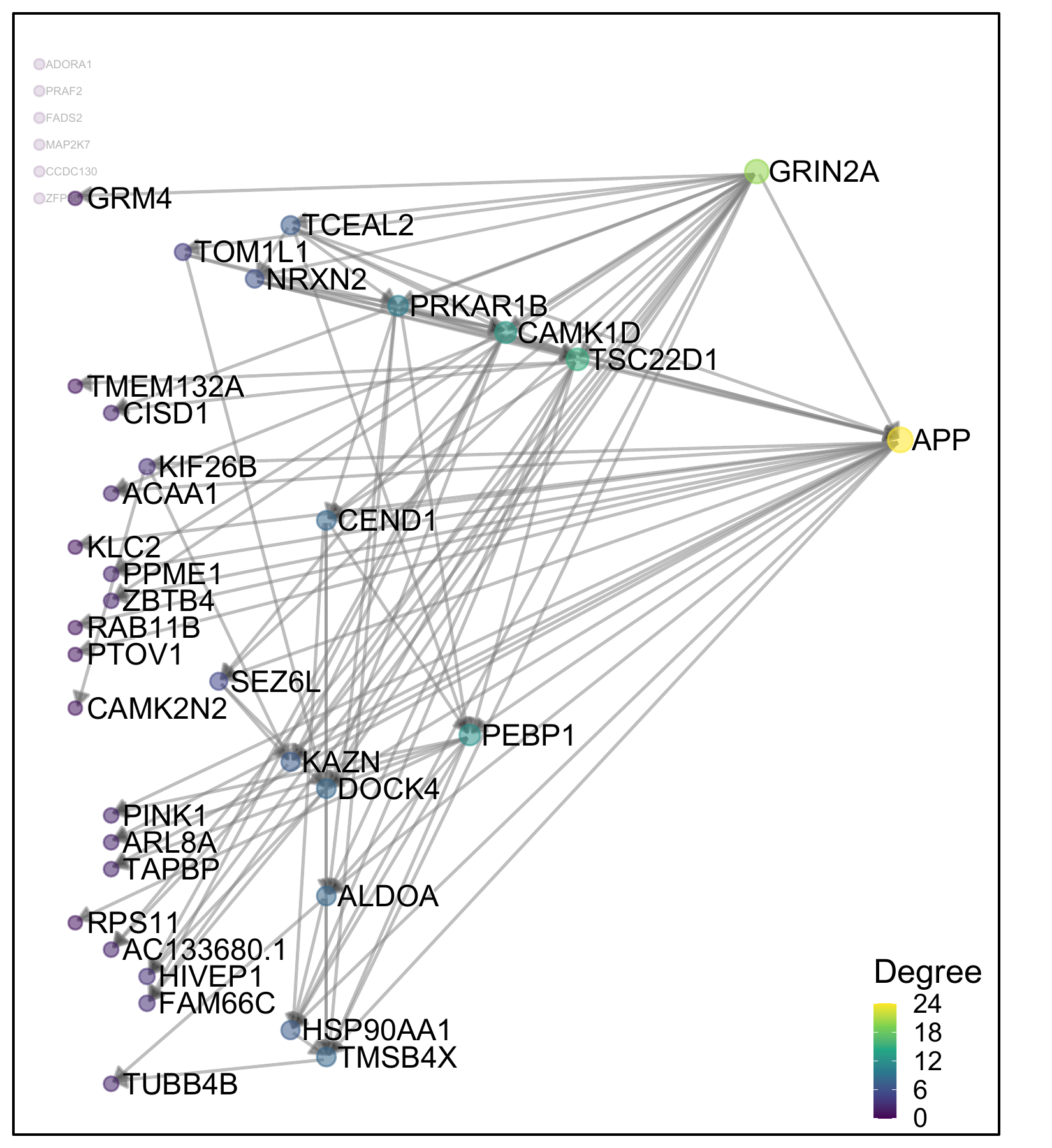}
    \includegraphics[width=0.49\linewidth]{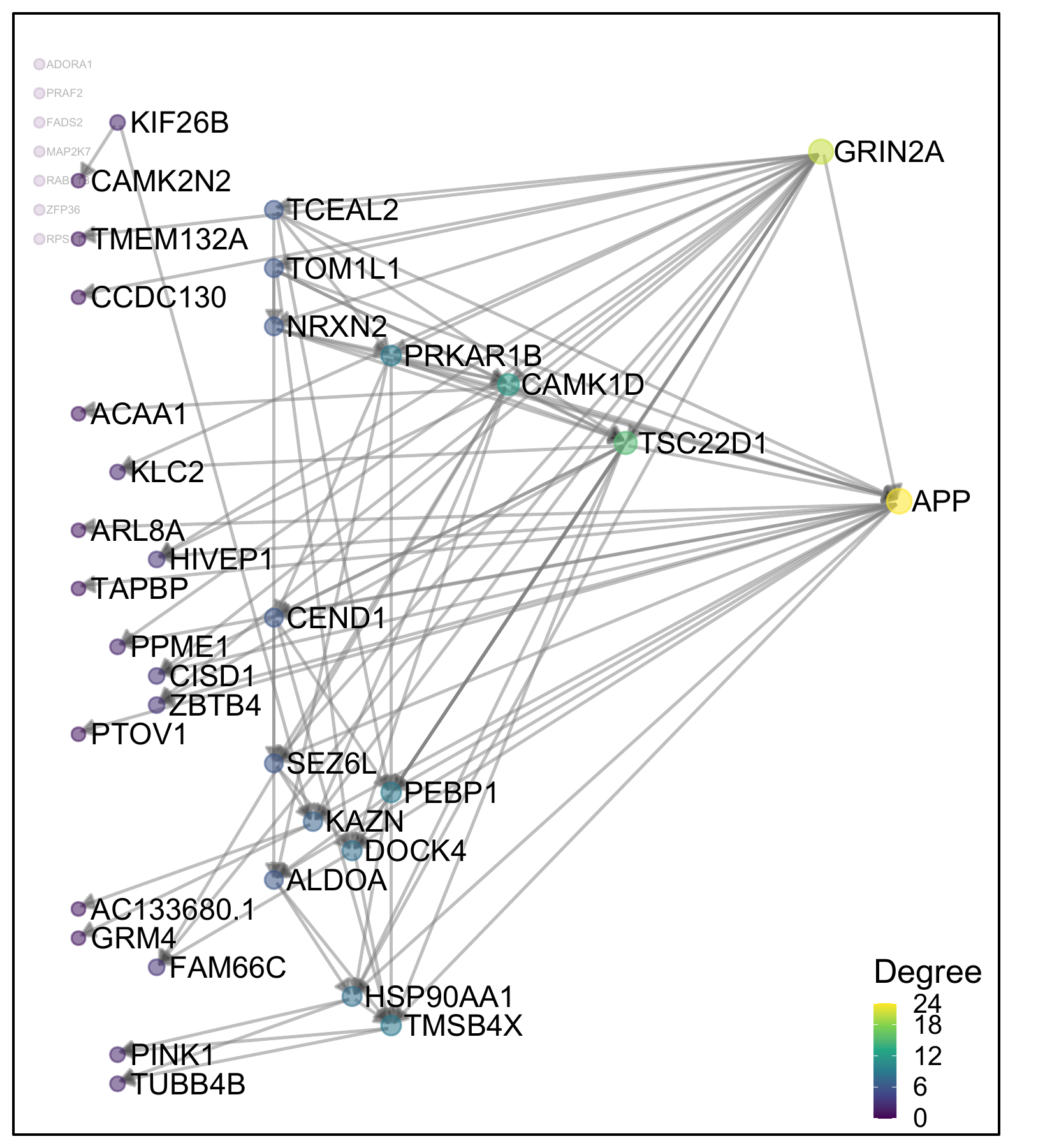}
    \caption{Consensus networks at voting threshold $\ell=8$: control (left), MDD (right). The networks summarize the subject-specific DAGs estimated using Algorithm~\ref{alg1}. Node color indicates total degree. Genes with no retained incoming or outgoing edges are shown faded.}
    \label{fig:real_data_group1}
\end{figure}

\begin{figure}[!t]
    \centering
    \captionsetup{font={normalsize,stretch=1}}

    \includegraphics[width=0.49\linewidth]{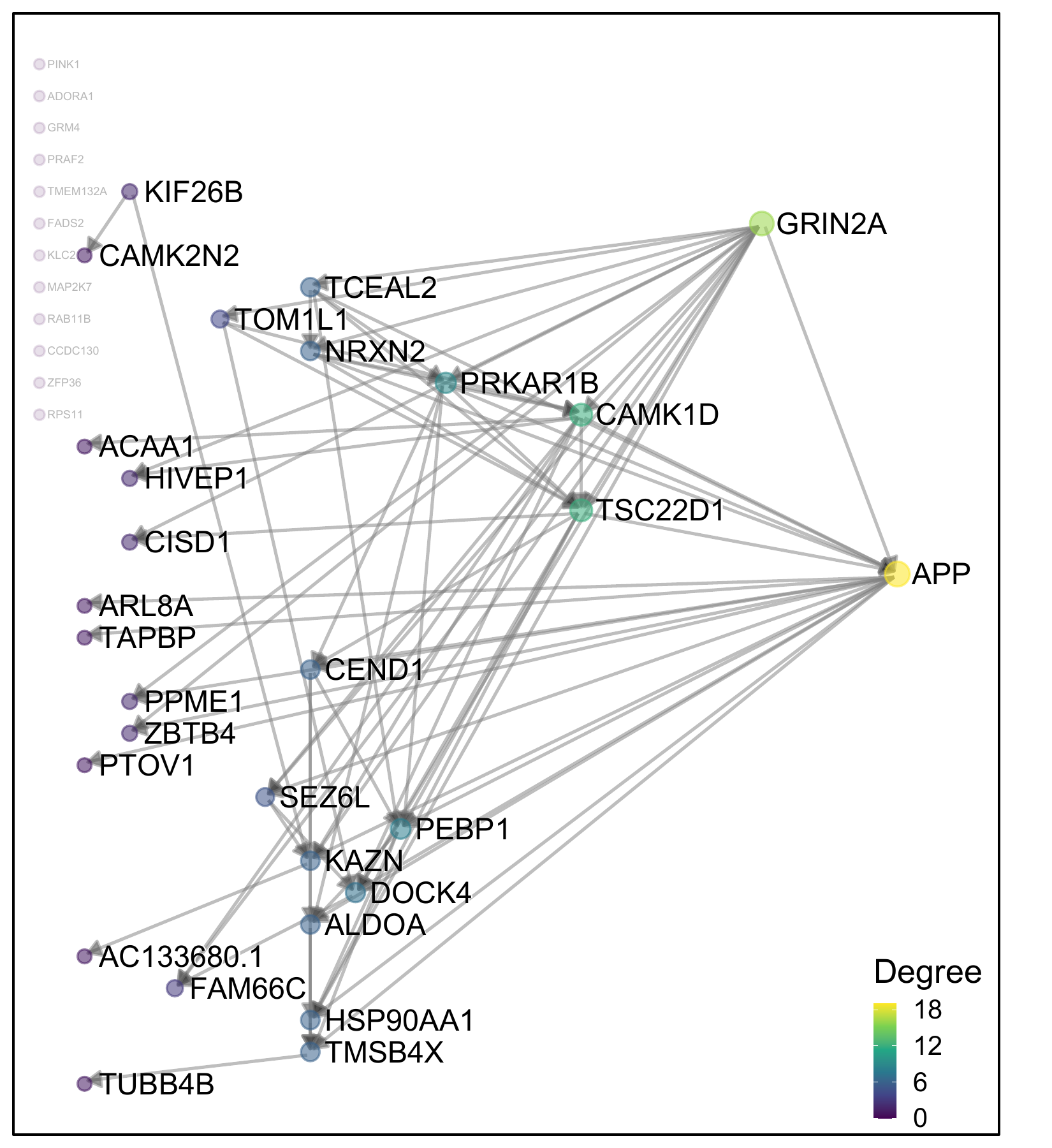}
    \includegraphics[width=0.49\linewidth]{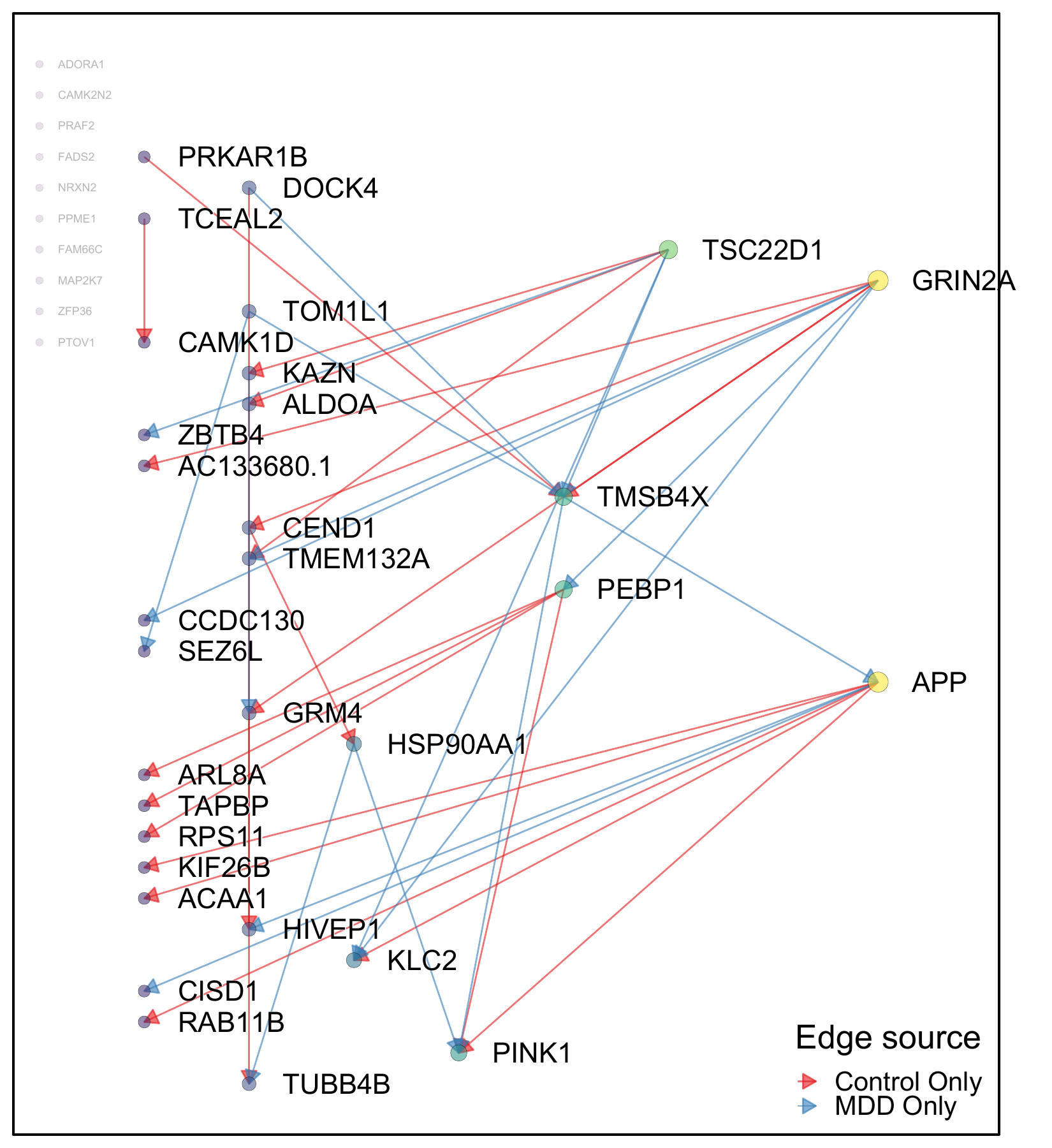}

    \caption{Consensus networks at voting threshold $\ell=8$, edges shared by the two groups (left), and edges retained in only one group (right).}
    \label{fig:real_data_group2}
    \vspace{-8pt}
\end{figure}

\textbf{Gene-level connectivity.} We further investigate how these differences are reflected in gene-level connectivity. To this end, we define weighted total connectivity by aggregating the selection frequencies of all incoming and outgoing edges. Figure~\ref{fig:weight_connectivity} compares weighted total connectivity between the control and MDD groups for the ten genes with the largest absolute differences.

\textbf{Consensus networks.}  
For network visualization, we construct group-specific consensus networks using a voting threshold of $\ell=8$, retaining edges selected in more than eight subjects within each group.
Figures~\ref{fig:real_data_group1} and~\ref{fig:real_data_group2} show the control and MDD networks and their intersection and differences, respectively. Sensitivity to the voting threshold is examined in the left panel of Figure~\ref{fig:edge_rank} and Section~\ref{subsec:voting_thres} of the supplementary material. Figure~\ref{fig:rank_plot} compares gene rankings by total degree (the number of incoming and outgoing edges) between the control and MDD consensus networks. Similarly, we also compare out-degree (the number of outgoing edges) rankings between the two groups in the supplementary material (right panel of Figure~\ref{fig:edge_rank}).

\section{Limitations}
\textbf{Scalability.} The proposed method can be computationally intensive when the number of nodes $p$ is large, since the order-search procedure is sequential and each candidate ordering requires repeated local parent-selection evaluations. Parallel computing across nodes and datasets, together with conditional-score shortlisting, partially mitigates this cost. With additional CPU resources, the procedure could be further accelerated using parallel randomized local search, which evaluates multiple neighbors concurrently at each iteration~\citep{neumann2007randomized, liu2000multiple,chang2022rapidly}. Nevertheless, the current implementation is intended for targeted modules containing tens of variables, and scaling to hundreds of variables requires further computational approximations (Supplementary Section~\ref{subsec:computational_cost}). 

\setcounter{figure}{3}
\renewcommand{\thefigure}{5.\arabic{figure}}

\begin{wrapfigure}[25]{r}{0.53\textwidth}
    \centering
    \captionsetup{font={small,stretch=1}}
    \includegraphics[width=.90\linewidth]{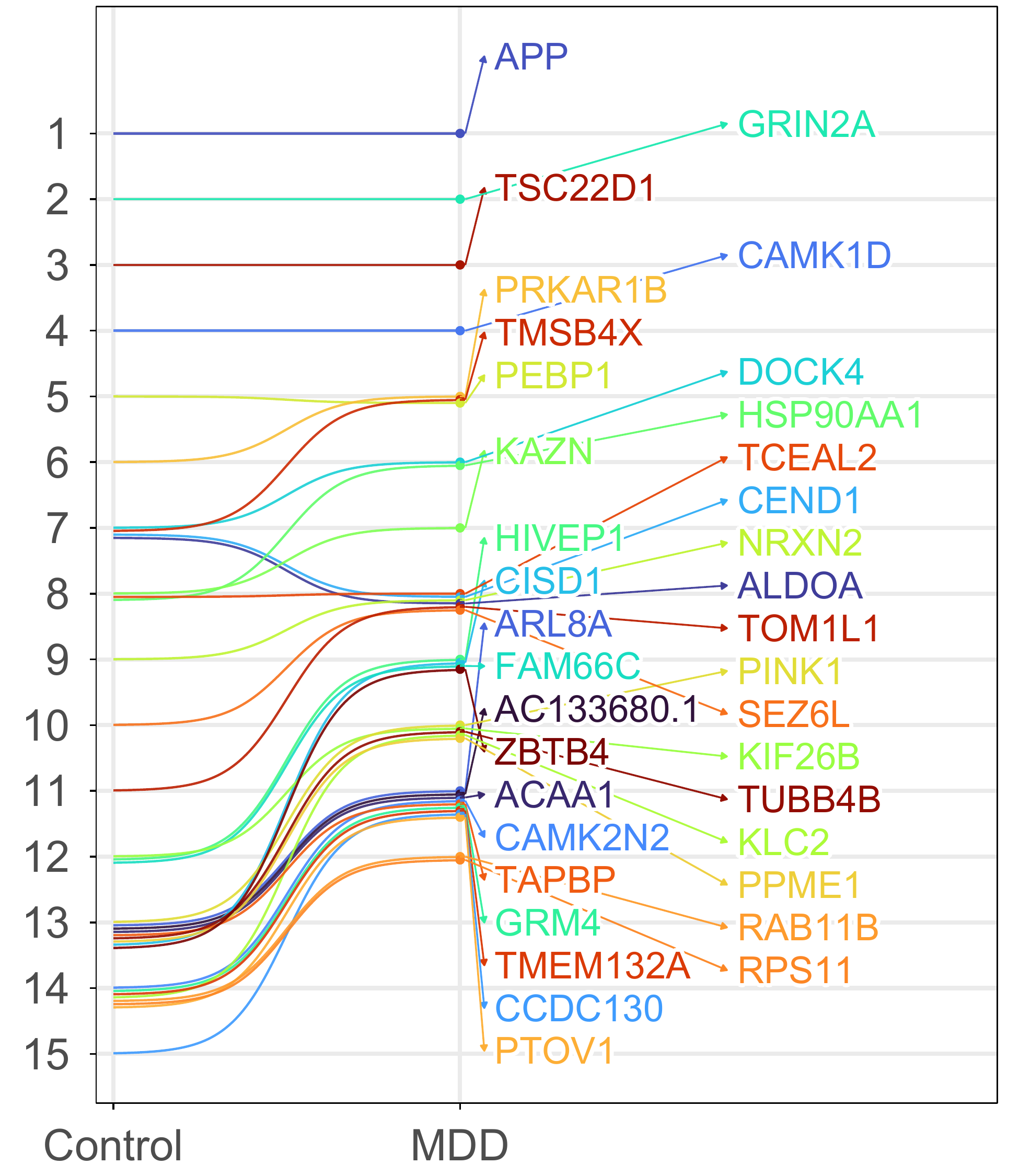}
    \caption{Total-degree rank plots (with rank = 1 indicating
    the highest connectivity) illustrate how the estimated
    network connectivity shifts between the control and MDD groups.}
    \label{fig:rank_plot}
\end{wrapfigure}

\renewcommand{\thefigure}{\thesection.\arabic{figure}}

\textbf{Common ordering assumption.} The proposed method assumes that all source-specific DAGs are consistent with at least one shared topological ordering. The assumption can be restrictive in settings where causal directions reverse across individuals, conditions, or biological subtypes. For example, if one subject has $i \to j$ while another has $j \to i$, then no single ordering can represent both graphs exactly. In such cases, the estimated ordering should be interpreted as a compromise ordering that captures dominant shared directional information. This limitation is especially relevant for systems with feedback-like behavior that is only approximately represented by a DAG. We conduct a sensitivity analysis in the supplementary material (Section~\ref{subsec:common}).

\textbf{Interpretation of real data networks.} Network estimates depend on the ZIP model and the common ordering assumption. Edge-frequency differences and degree rankings are descriptive, and establishing disease-associated causal changes requires further biological validation. We provide a diagnostic assessment of the common ordering assumption in the supplementary material (Section~\ref{subsec:shared_order}).


\newpage
\begingroup
\begin{ack}
Chang's research was supported in part by faculty start-up funds from the University of Texas at Dallas. 
Choi's research was supported in part by the National Research Foundation of Korea (NRF) grant funded by the Korean government (MSIT) (No. RS-2022-NR068754) and by the Institute of Information \& Communications Technology Planning \& Evaluation (IITP) grant funded by the Korean government (MSIT) (No. RS-2026-25507543).

\end{ack}

\small
\bibliographystyle{plainnat}
\bibliography{reference}
\endgroup
\clearpage
\appendix
\section*{Supplementary material for ``ZIO: Order-Based Structure Learning for Zero-Inflated Count Data under Data Heterogeneity''}

\setcounter{section}{0}
\setcounter{subsection}{0}
\setcounter{figure}{0}
\setcounter{table}{0}
\setcounter{equation}{0}
\renewcommand{\thesection}{\Alph{section}}
\renewcommand{\thesubsection}{\thesection\arabic{subsection}}
\renewcommand{\thefigure}{\thesection\arabic{figure}}
\renewcommand{\thetable}{\thesection\arabic{table}}
\renewcommand{\theequation}{\thesection\arabic{equation}}

\section{Proofs}\label{sec:proofs}

\subsection{Proof of Theorem~\ref{thm:consistency}}\label{subsec:thm1}


\begin{proof}
For a graph \(G\in\mathcal G_p\), define the graph-level penalized score
\[
Q_n(G)
=
2\ell_{G,n}(\widehat\Theta_G)-c_0d_G\log n,
\qquad
d_G=2|E(G)|+2p.
\]
Since both the likelihood and the penalty decompose node-wise, for every ordering
\(\sigma\),
\[
\mathrm{OBS}(\sigma)
=
\max_{G\in\mathcal G_p^\sigma} Q_n(G).
\]

We first show that, with probability tending to one, \(G^*\) has larger
penalized score than every \(G\neq G^*\).

Fix \(G\in\mathcal G_p\).

Let $\widetilde\ell_G(\Theta_G)=\mathbb{E}_{p^*}\log p_G(\mathbf X;\Theta_G)$ denote the expected log-likelihood under $G$, and let $m_G=\sup_{\Theta_G\in\Omega_G}\widetilde\ell_G(\Theta_G)$ denote its supremum over $\Omega_G$.
Under Assumption~\ref{assump:lik_regular}, this supremum is
attained at some population maximizer $\widetilde\Theta_G\in\Omega_G$, and
the uniform law of large numbers yields
$\sup_{\Theta_G\in\Omega_G}\bigl|n^{-1}\ell_{G,n}(\Theta_G)
-\widetilde\ell_G(\Theta_G)\bigr|=o_p(1)$. Since $\widehat\Theta_G$ maximizes
$\ell_{G,n}$ over $\Omega_G$, it follows that
\begin{equation}
n^{-1}\ell_{G,n}(\widehat\Theta_G)=m_G+o_p(1).
\label{eq:pop-expansion-G}
\end{equation}

We next identify $m_{G^*}$. Since $p^*=p_{G^*}(\cdot;\Theta^*)$, for any
$\Theta\in\Omega_{G^*}$,
\[
\widetilde\ell_{G^*}(\Theta^*)-\widetilde\ell_{G^*}(\Theta)
=\mathrm{KL}\!\left(p^*\,\middle\|\,p_{G^*}(\cdot;\Theta)\right)\ge 0.
\]

Thus the true parameter $\Theta^*$ is a population maximizer under $G^*$, implying that
$m_{G^*}=\widetilde\ell_{G^*}(\Theta^*)=\mathbb{E}_{p^*}\log p^*(X)$.
Applying \eqref{eq:pop-expansion-G} with $G=G^*$ gives
\begin{equation}
n^{-1}\ell_{G^*,n}(\widehat\Theta_{G^*})
=\mathbb{E}_{p^*}\log p^*(X)+o_p(1).
\label{eq:pop-expansion-Gstar}
\end{equation}
\color{black}

Now consider first a graph \(G\not\supseteq G^*\). Combining
\eqref{eq:pop-expansion-G} and \eqref{eq:pop-expansion-Gstar},
\[
\begin{aligned}
Q_n(G^*)-Q_n(G)
&=
2\left\{
\ell_{G^*,n}(\widehat\Theta_{G^*})
-
\ell_{G,n}(\widehat\Theta_G)
\right\}
-
c_0(d_{G^*}-d_G)\log n  \\
&=
2n\left\{
\mathbb{E}_{p^*}\log p^*(X)
-
m_G
\right\}
-
c_0(d_{G^*}-d_G)\log n
+
o_p(n).
\end{aligned}
\]

The coefficient of $2n$ is the smallest Kullback--Leibler divergence between
$p^*$ and the model under $G$:
\[
\mathbb{E}_{p^*} \log p^*(\mathbf X) - m_G
=
\min_{\Theta_G\in\Omega_G}
\mathrm{KL}
\!\left(
p^*
\,\middle\|\,
p_G(\cdot;\Theta_G)
\right)
= 
\mathrm{KL}\!\left(p^*\,\middle\|\,p_G(\cdot;\widetilde\Theta_G)\right).
\]
This divergence is strictly positive whenever $G\not\supseteq G^*$. Indeed, if it were zero, $p^*$ would be representable by the ZIPBN model on $G$, which contradicts the identifiability of ZIPBN models (Theorem~1 of \citealp{choi2020bayesian}) and the causal minimality condition in Assumption~\ref{assump:minimal}. 
\color{black}

Consequently, for every such $G$,
\[
Q_n(G^*)-Q_n(G)
=
2n\,\mathrm{KL}\!\left(p^*\,\middle\|\,p_G(\cdot;\widetilde\Theta_G)\right)
+
O(\log n)
+
o_p(n),
\]
and the leading term is positive and of order \(n\), whereas the penalty
difference is only of order \(\log n\). Hence,
\[
Q_n(G^*)-Q_n(G)\xrightarrow{p}+\infty
\qquad
\text{for every }G\not\supseteq G^*.
\]

Next consider \(G\supsetneq G^*\). By
Assumption~\ref{assump:lik_regular}, the model under \(G^*\)
is a regular nested submodel of the model associated with \(G\), and
Wilks' theorem applies to the restrictions corresponding to the extra
edges in \(E(G)\setminus E(G^*)\). Hence
\[
2\left\{
\ell_{G,n}(\widehat\Theta_G)
-
\ell_{G^*,n}(\widehat\Theta_{G^*})
\right\}
\stackrel{d}{\longrightarrow}
\chi^2_{d_G-d_{G^*}}.
\]
In particular,
\[
2\left\{
\ell_{G,n}(\widehat\Theta_G)
-
\ell_{G^*,n}(\widehat\Theta_{G^*})
\right\}
=O_p(1).
\]
Since \(d_G>d_{G^*}\), we have
\[
\begin{aligned}
Q_n(G^*)-Q_n(G)
&=
-2\left\{
\ell_{G,n}(\widehat\Theta_G)
-
\ell_{G^*,n}(\widehat\Theta_{G^*})
\right\}
+
c_0(d_G-d_{G^*})\log n \\
&\xrightarrow{p}+\infty.
\end{aligned}
\]

Because \(p\) is fixed, \(\mathcal G_p\) is finite. Hence the preceding
pointwise comparisons hold uniformly over all \(G\neq G^*\). Equivalently,
\[
\Pr\left\{
Q_n(G^*)>Q_n(G)
\text{ for all }G\neq G^*
\right\}
\longrightarrow 1.
\]
Denote this high-probability event by \(\mathcal E_n\).

We now prove the ordering statement. Let \(\tau\notin\mathcal L(G^*)\). Then
no graph \(G\in\mathcal G_p^\tau\) can contain \(G^*\); otherwise \(\tau\)
would respect every edge of \(G^*\) and hence would be a topological ordering
of \(G^*\), contradicting \(\tau\notin\mathcal L(G^*)\). Therefore, on
\(\mathcal E_n\),
\[
Q_n(G^*)
>
\max_{G\in\mathcal G_p^\tau} Q_n(G)
=
\mathrm{OBS}(\tau).
\]
On the other hand, if \(\sigma\in\mathcal L(G^*)\), then
\(G^*\in\mathcal G_p^\sigma\), and thus
\[
\mathrm{OBS}(\sigma)
=
\max_{G\in\mathcal G_p^\sigma} Q_n(G)
\ge Q_n(G^*).
\]
Consequently, on \(\mathcal E_n\),
\[
\mathrm{OBS}(\sigma)
>
\mathrm{OBS}(\tau)
\qquad
\text{for every }
\sigma\in\mathcal L(G^*),\ 
\tau\notin\mathcal L(G^*).
\]
Since \(\Pr(\mathcal E_n)\to 1\), this proves part~(1).

Finally, fix \(\sigma\in\mathcal L(G^*)\). Since \(G^*\in\mathcal G_p^\sigma\)
and, on \(\mathcal E_n\),
\[
Q_n(G^*)>Q_n(G)
\qquad
\text{for all }
G\in\mathcal G_p^\sigma\setminus\{G^*\},
\]
the maximizer of \(Q_n(G)\) over \(\mathcal G_p^\sigma\) is uniquely \(G^*\).
Hence
\[
\widehat G_\sigma
=
\arg\max_{G\in\mathcal G_p^\sigma} Q_n(G)
=
G^*
\]
with probability tending to one. This proves part~(2).
\end{proof}

\subsection{Proof of Corollary~\ref{cor:multi}}\label{subsec:cor1}

Again, recall that $G^{(k)}$ denotes the true DAG for the dataset $\bX^{(k)}$ and $\Theta^{(k)}$ the corresponding true parameters. The joint order-based score is defined by
\begin{align*}
   \mathrm{JOBS}(\sigma) = \sum_{k=1}^K \mathrm{OBS}^{(k)}(\sigma).
\end{align*}

\begin{proof}
Fix
\[
\sigma\in\bigcap_{k=1}^K\mathcal L(G^{(k)})
\]
and let
\[
\tau\notin\bigcap_{k=1}^K\mathcal L(G^{(k)}).
\]
Then there exists at least one index \(k\) such that
\(\tau\notin\mathcal L(G^{(k)})\). For every such \(k\), Theorem~\ref{thm:consistency}
applied to the \(k\)-th dataset gives
\[
\mathbb P\left\{
\mathrm{OBS}^{(k)}(\sigma)>
\mathrm{OBS}^{(k)}(\tau)
\right\}\to 1 .
\]
For any \(k'\) such that \(\tau\in\mathcal L(G^{(k')})\), both \(\sigma\) and
\(\tau\) are consistent orderings of \(G^{(k')}\). Hence, by the graph-recovery part
of Theorem~\ref{thm:consistency},
\[
\mathbb P\left\{
\widehat G^{(k')}_\sigma=G^{(k')}
\text{ and }
\widehat G^{(k')}_\tau=G^{(k')}
\right\}\to 1,
\]
which implies
\[
\mathrm{OBS}^{(k')}(\sigma)=\mathrm{OBS}^{(k')}(\tau)
\]
with probability tending to one. Since \(K\) is fixed, combining these statements over \(k=1,\ldots,K\) by a
finite union bound yields
\[
\mathbb P\left\{
\mathrm{JOBS}(\sigma)>\mathrm{JOBS}(\tau)
\right\}\to 1 .
\]

Finally, since \(\sigma\in\mathcal L(G^{(k)})\) for every \(k\), another
application of Theorem~\ref{thm:consistency} gives
\[
\mathbb P\left\{
\widehat G^{(k)}_\sigma=G^{(k)}
\right\}\to 1
\qquad\text{for each }k\in[K].
\]
Again using that \(K\) is fixed,
\[
\mathbb P\left\{
\bigcap_{k=1}^K
\{\widehat G^{(k)}_\sigma=G^{(k)}\}
\right\}
\ge
1-\sum_{k=1}^K
\mathbb P\{\widehat G^{(k)}_\sigma\neq G^{(k)}\}
\to 1
\]
as \(\min_{k\in[K]} n_k\to\infty\). This completes the proof.

\end{proof}

\subsection{Model-specific sufficient conditions for Assumption~\ref{assump:lik_regular}}\label{subsec:alt_assump2}

We provide a set of model-specific likelihood regularity conditions that are sufficient for Assumption 2. These conditions are more intuitive than Assumption 2 itself, albeit stronger. For a node $j$ and a candidate parent set $S$, let $\Omega_{j, S}$ be the node-wise parameter space, let $\tilde{\ell}_{j, S}(\theta_{j, S}) = E_{p^*} \{ \ell_j(S, \theta_{j, S}; \bm{X}_j, \bm{X}_S) \}$ denote the node-wise expected log-likelihood, and let $\tilde{\theta}_{j, S} \in \arg\max_{\theta_{j, S} \in \Omega_{j, S}} \tilde{\ell}_{j, S}(\theta_{j, S})$ be one of its maximizers.

\begin{enumerate}[label=(R\arabic*), ref=(R\arabic*)]
\item\label{cond:compact} \textbf{Compact interior parameterization.}
For each $j$ and $S$, the parameter space is a compact set
\[
    \Omega_{j, S}=[-M_{j},M_{j}]^{2|S|+2}
\]
for some $0 < M_{j} < \infty$, and $\tilde{\theta}_{j, S}$ lies in the interior of $\Omega_{j, S}$. Moreover, for any overfitted graph $G \supseteq G^*$,
the true parameter $\theta_{j, \Pa_j(G)}^*$ of node $j$, padded with $(0, 0)$ in the coordinates of the extra parents $\Pa_j(G) \setminus \Pa_j(G^*)$, is also an interior point of $\Omega_{j,\Pa_j(G)}$.

\item\label{cond:nondeg} \textbf{Nondegenerate ZIP design.}
At the true parameter,
\[
    0<\eta_j^*(x_{\operatorname{Pa}_j(G^*)})<1,
    \qquad
    \lambda_j^*(x_{\operatorname{Pa}_j(G^*)}) > 0
\]
for every vector of parent counts $x_{\operatorname{Pa}_j(G^*)}$ in the support.

\item\label{cond:moment} \textbf{Exponential moments.}
For every \(t>0\),
\[
    \mathbb{E}_{p^*}
    \left[
        \exp\left\{t \sum_{j=1}^p X_j \right\}
    \right]
    <\infty.
\]
\end{enumerate}

Under \ref{cond:compact}--\ref{cond:moment}, the standard asymptotic theory for the MLE applies to the ZIP model; the argument rests on three ingredients. 
First, continuity of the ZIP log-likelihood and compactness of the parameter spaces \ref{cond:compact} ensure the existence of a global MLE. Finite-sample uniqueness is not required, and any element of the argmax may be selected. 
Second, the exponential-moment condition \ref{cond:moment} supplies integrable envelopes for the log-likelihood and its first three derivatives, so that the likelihood classes are $p^*$-Glivenko–Cantelli. This yields the required uniform law of large numbers and the local likelihood expansion. A simple sufficient condition for \ref{cond:moment} is that the true conditional Poisson rates $\lambda_j^*(x_{\operatorname{Pa}_j(G^*)})$ be uniformly bounded over the support of the parent counts, ruling out an explosion of counts. 
Third, the nondegenerate design condition \ref{cond:nondeg}, $0<\eta_j^*<1$ and $\lambda_j^*>0$, ensures that the joint distribution has full support on $\mathbb{N}_0^p$, so that no nonzero linear combination of $Z_S = (X_S^\top, 1)^\top$ vanishes almost surely; combined with the finite second moments implied by \ref{cond:moment}, this yields $\mathbb{E}_{p^*}\!\left[ Z_S Z_S^\top \right] \succ 0$, and hence identifiability of the regression coefficients and nonsingularity of the Fisher information. 
Together, these ingredients establish the standard asymptotic results for the ZIP model and suffice to prove Theorem~1.

For related likelihood theory, see \citet{czado2005consistency} and \citet{czado2007zero}, who establish consistency and asymptotic normality for zero-inflated generalized Poisson regression, which contains ZIP regression as a special case, under interiority and information-regularity conditions. Because our regressors are random, unbounded counts, their bounded-design condition does not apply; we replace it with the exponential-moment condition, which provides the required likelihood-derivative envelopes. The remaining M-estimation and Wilks arguments follow standard likelihood theory \citep{van2000asymptotic}. 

\color{black}

\newpage

\section{Additional simulation studies}

\subsection{Simulation settings and implementation details}
\label{subsec:simulation_settings}

\subsubsection{Data generating parameters}
\label{subsec:zero_inflation_ranges}
In Section~\ref{subsec:zero_rate}, we evaluate the sensitivity of the proposed method to the overall level of zero inflation. To obtain simulated datasets with $25\%$, $50\%$, and $75\%$ zeros, we follow the parameter ranges used by \cite{choi2020bayesian}. The $50\%$ zero setting is the default setting used in the main simulation study in Section~\ref{subsec:performance}. For each zero proportion setting, the parameters in the zero-inflation and Poisson components are generated from the ranges shown in Table~\ref{tab:zero_parameter_ranges}.
\begin{table}[!htb]
\captionsetup{font={normalsize,stretch=1}}
\caption{Parameter ranges used to generate datasets with approximately $25\%$, $50\%$, and $75\%$ zeros.}
\label{tab:zero_parameter_ranges}
\centering
\small
\renewcommand{\arraystretch}{1.3}
\begin{tabular}{ccccc}
\toprule
Target zero proportion 
& $\alpha_{ij}^{(k)}$ 
& $\beta_{ij}^{(k)}$ 
& $\delta_j^{(k)}$ 
& $\gamma_j^{(k)}$ \\
\midrule
$25\%$ 
& $\mathrm{Unif}(-1.0,-0.3)$ 
& $\mathrm{Unif}(-0.4,-0.2)$ 
& $\mathrm{Unif}(-2.0,-1.5)$ 
& $\mathrm{Unif}(1.5,2.0)$ \\

$50\%$ 
& $\mathrm{Unif}(0.3,1.0)$ 
& $\mathrm{Unif}(-1.0,-0.3)$ 
& $\mathrm{Unif}(-2.0,-1.0)$ 
& $\mathrm{Unif}(1.0,2.0)$ \\

$75\%$ 
& $\mathrm{Unif}(0.3,1.0)$ 
& $\mathrm{Unif}(-1.0,-0.3)$ 
& $\mathrm{Unif}(0.5,1.0)$ 
& $\mathrm{Unif}(1.0,1.5)$ \\
\bottomrule
\end{tabular}
\end{table}

\begin{table}[!b]
\caption{Hyperparameter and edge-selection settings used for the five competing methods in the method comparison study. Values listed for $p=(25,50,75)$ are given in that order.}
\label{tab:method_hyperparameters}
\centering
\small
\renewcommand{\arraystretch}{1.3}
\begin{tabular}{lll}
\toprule
Method & Hyperparameter or setting & Value \\
\midrule
\multirow{7}{*}{ZIPBN}
& Number of MCMC iterations & $3000$ \\
& Burn-in & $1500$ \\
& Number of chains & $10$ \\
& Swap probability & $0.1$ \\
& Temperature ladder & $\exp\{\mathrm{seq}(0,1,\mathrm{length}=10)\}$ \\
& Prior for $\tau$ & $(0.01,0.01)$ \\
& Prior for $\rho$ & $(0.5,0.5)$ \\
\midrule
ODS
& Minimum conditioning-set proportion & $c_0=0.004$ \\
\midrule
MRS
& Minimum cell count & $N_{\min}=2$ \\
\midrule
\multirow{6}{*}{ZICO}
& Loss type & Poisson \\
& Maximum iterations & $3000$ \\
& \texttt{lam} & $10^{-3}$ \\
& \texttt{lambda\_align} & $1$ \\
& Edge score & $|W_0|+|W_1|$ \\
& Edge threshold & $0.30$ \\
\midrule
\multirow{6}{*}{NOTEARS}
& Loss type & Poisson \\
& Maximum iterations & $100$ \\
& Acyclicity tolerance $h_{\mathrm{tol}}$ & $10^{-8}$ \\
& Maximum penalty $\rho_{\max}$ & $10^{16}$ \\
& $\ell_1$ penalty $\lambda_1$ for $p=(25,50,75)$
    & $(0.0009,\,0.0009,\,0.0020)$ \\
& Edge threshold for $p=(25,50,75)$
    & $(0.30,\,0.40,\,0.30)$ \\
\bottomrule
\end{tabular}
\end{table}

\subsubsection{Settings for competing methods}
\label{subsec:method_hyperparameters}
{Table~\ref{tab:method_hyperparameters} summarizes the hyperparameter settings for the methods compared in Section~\ref{subsec:performance}. For ZIPBN, ODS, and MRS, we follow \cite{choi2020bayesian}. For ZICO and NOTEARS, we use the authors' original implementations \citep{sato2025dag,zheng2018dags} and tune the hyperparameters for our simulation settings.

\subsection{Validation of the proposed estimation procedure}
\label{subsec:optimization}

\subsubsection{Score separation and graph recovery}
\label{subsec:theoretical_results}

\textbf{Score separation.} For a fixed true DAG $G^*$, we consider a collection of consistent topological orderings, $\mathcal{L}(G^*)$, and a collection of inconsistent orderings, $\mathcal{L}(G^*)^c$. For each ordering, we compute the order-based scores $\mathrm{OBS}(\sigma)$ and $\mathrm{OBS}(\tau)$, and compare their behavior across repeated simulations.

We consider $p \in \{5,10\}$, $K=1$, $n \in \{250,500,1000,2000\}$, and a zero proportion of approximately $50\%$, following the simulation setup in Section~\ref{subsec:performance}.  
Let $\mathcal{V}_{\mathrm{eval}} \subseteq \mathcal{L}(G^*)$
and $\mathcal{I}_{\mathrm{eval}} \subseteq \mathcal{L}(G^*)^c$
denote the collections of evaluated consistent and inconsistent orderings, respectively. For $p=5$, we enumerate all orderings in both categories, whereas for $p=10$, we randomly sample 30 from each. We quantify their separation using the
score gap
\[
\widehat{\mathrm{gap}}_n
=
\min_{\sigma \in \mathcal{V}_{\mathrm{eval}}}
\mathrm{OBS}(\sigma)
-
\max_{\tau \in \mathcal{I}_{\mathrm{eval}}}
\mathrm{OBS}(\tau).
\]
A positive value of $\widehat{\mathrm{gap}}_n$ indicates that all evaluated consistent orderings receive better objective scores than the evaluated inconsistent orderings.
\begin{figure}[h!]
\centering
\captionsetup{font={normalsize,stretch=1}}
\scalebox{1}{ 
    \begin{minipage}{0.65\textwidth}  
            \includegraphics[width=\linewidth]{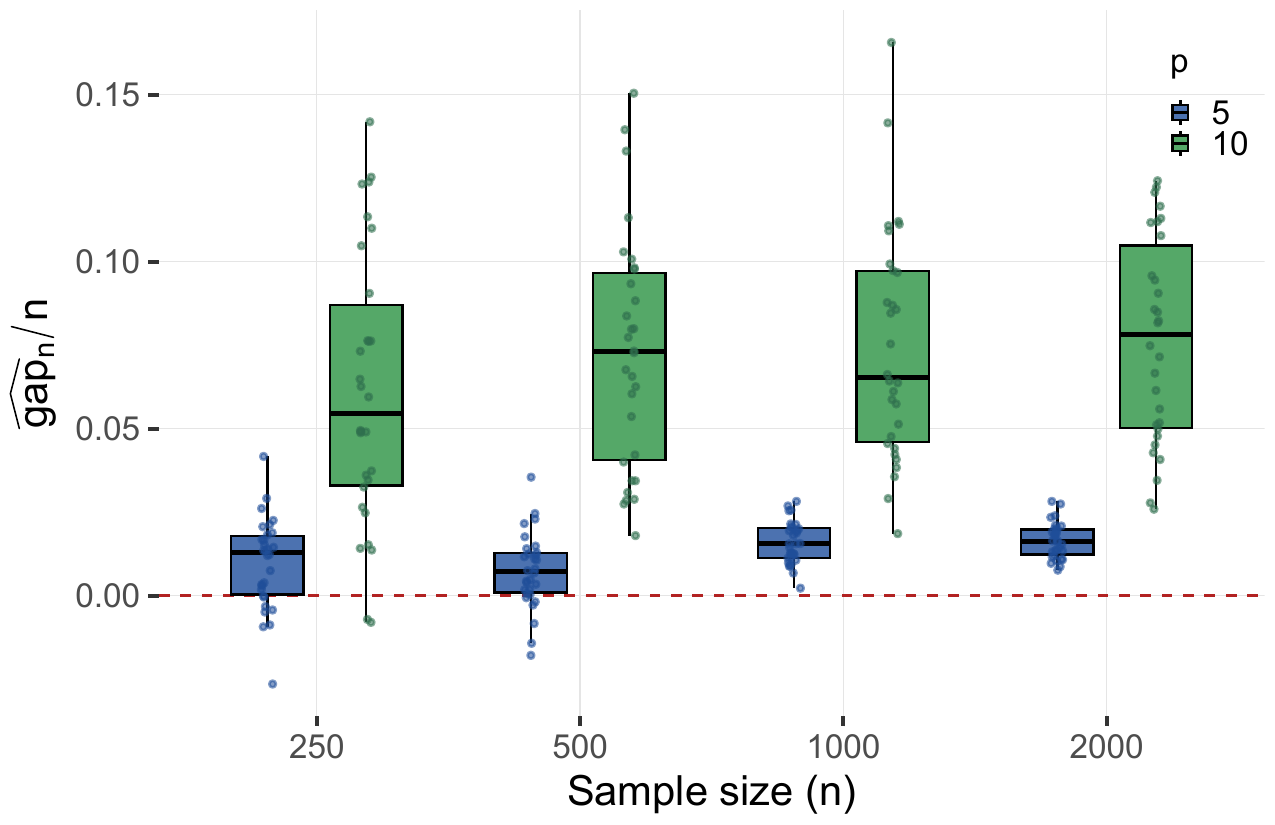}  
    \end{minipage}%
}
    \captionof{figure}{Boxplots of the scaled score gap $\widehat{\mathrm{gap}}_n/n$ across repeated simulations for different sample sizes $n$ and numbers of nodes $p$. The dashed red line indicates zero; positive values indicate separation in favor of consistent orderings.}
        \label{fig:theorem_score_gap}
\end{figure}
Figure~\ref{fig:theorem_score_gap} shows that the scaled score gap is generally positive across all sample sizes and both values of $p$, indicating that consistent orderings tend to receive higher objective scores than inconsistent orderings. The separation is more pronounced for $p=10$, while for $p=5$ the gaps are smaller but mostly remain above zero. This provides finite-sample support for the score-separation behavior described in Theorem~\ref{thm:consistency}.
\begin{figure}[t!]
\centering
\captionsetup{font={normalsize,stretch=1}}
\scalebox{1}{ 
    \begin{minipage}{0.65\textwidth}  
            \includegraphics[width=\linewidth]{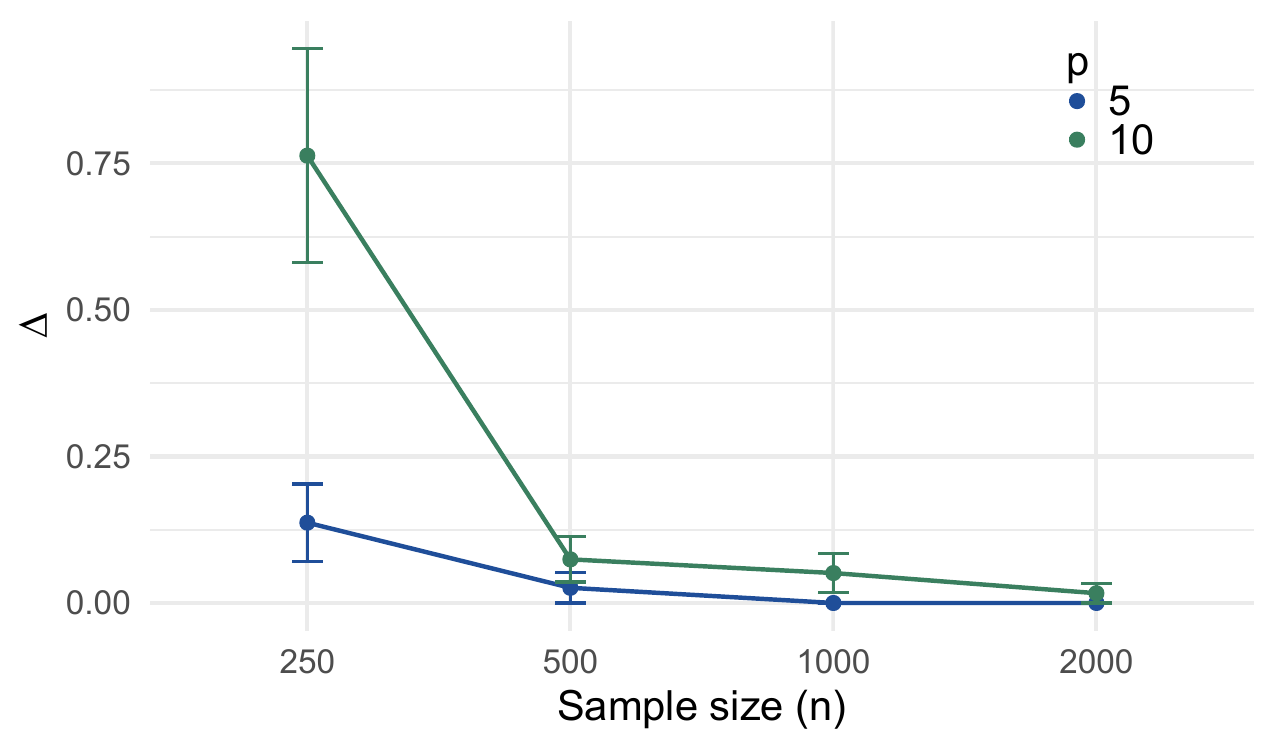}  
    \end{minipage}%
}
\caption{Mean structural Hamming distance, $\Delta$, with standard errors as a function of sample size $n$, restricted to consistent topological orderings. The blue line corresponds to $p=5$, and the green line corresponds to $p=10$.}
\label{fig:theorem_valid_order_shd}
\end{figure}

\textbf{Graph recovery.} We evaluate graph recovery under consistent orderings using the structural Hamming distance \(\Delta\), with standard errors across repeated simulations. Figure~\ref{fig:theorem_valid_order_shd} shows that recovery error decreases as \(n\) increases and remains close to zero after the initial decrease from \(n=250\) to \(n=500\). Thus, the experiment supports both components of the theorem: separation of consistent and inconsistent orderings and increasingly accurate graph recovery when a consistent ordering is available.

\subsubsection{Accuracy of approximate parent selection} \label{subsec:approc_parent}
To assess the effect of approximate parent selection, we compare exact optimization with the forward--backward stepwise procedure. 
We generate sparse ZIP DAGs with \(p=7\) following~\cite{choi2020bayesian}. For each dataset, we enumerate all \(7! = 5{,}040\) orderings, and compute the order-based scores using both exact and forward–backward parent selection. We consider \(n\in\{250,500,1000\}\), with \(100\)  simulation replicates per sample size.  Table~\ref{tab:exact-stepwise-ablation} reports three measures. \textbf{Agreement} is the proportion of node–ordering pairs for which exhaustive and stepwise optimization select identical parent sets. \textbf{Optimal recovery} is the percentage of replicates in which at least one globally BIC-optimal ordering is a consistent topological ordering of the generating DAG. \textbf{SHD} measures the structural difference between the true DAG and the DAG corresponding to a globally optimal ordering.
As an additional measure of global-optimum overlap, the exact and stepwise objectives share at least one globally optimal ordering in \(99\%\), \(98\%\), and \(100\%\) of replicates for \(n\in\{250,500,1000\}\), respectively. Across the \(300\) evaluations, there are only two cases in which exact optimization identifies a true-compatible global optimum but the stepwise approximation does not. 

\begin{table}[!h]
\centering
\caption{Comparison of exact and forward--backward parent-set optimization in the exhaustive \(p=7\) ZIP DAG experiment. Values after \(\pm\) are standard errors across \(100\) replicates.}
\label{tab:exact-stepwise-ablation}
\small
\begin{tabular}{rccc}
\toprule
\(n\)
&
\begin{tabular}[c]{@{}c@{}}
\textbf{Agreement} (\%)
\end{tabular}
&
\begin{tabular}[c]{@{}c@{}}
\textbf{Optimal recovery}\\
Exact / Stepwise (\%)
\end{tabular}
&
\begin{tabular}[c]{@{}c@{}}
\textbf{SHD}\\
Exact / Stepwise
\end{tabular}
\\
\midrule
250
& \(99.45 \pm 0.11\)
& \(68.0 / 68.0\)
& \(1.23 \pm 0.16 \,/\, 1.26 \pm 0.16\)
\\
500
& \(99.62 \pm 0.11\)
& \(86.0 / 84.0\)
& \(0.51 \pm 0.11 \,/\, 0.59 \pm 0.12\)
\\
1000
& \(99.81 \pm 0.07\)
& \(94.0 / 94.0\)
& \(0.15 \pm 0.05 \,/\, 0.14 \pm 0.04\)
\\
\bottomrule
\end{tabular}
\end{table}

\begin{table}[!b]
\centering
\captionsetup{font={normalsize,stretch=1}}
\caption{R2R local-mode frequencies, global-basin coverage, and finite-search success rates (\%) for the exhaustive $p=7$ experiment.}
\label{tab:r2r_search_diagnostics}
\small
\begingroup
\renewcommand{\arraystretch}{1.3}
\begin{tabular}{ccccccc}
\toprule
& \multicolumn{2}{c}{Local-mode frequency}
& \multicolumn{2}{c}{Global-basin coverage}
& \multicolumn{2}{c}{Finite-search success} \\
\cmidrule(lr){2-3}
\cmidrule(lr){4-5}
\cmidrule(lr){6-7}
$n$ & Exact & Stepwise & Exact & Stepwise & Single start & Five starts \\
\midrule
250  & 47 & 46 & 84.2 & 84.9 & 76.3 & 95.5 \\
500  & 38 & 39 & 88.4 & 88.3 & 80.6 & 97.6 \\
1000 & 40 & 41 & 89.9 & 90.1 & 82.2 & 98.6 \\
\bottomrule
\end{tabular}%
\endgroup
\end{table}

\subsubsection{Order-search landscape and multiple initializations}
Using the exhaustive $p=7$ exact-versus-stepwise experiment in Section~\ref{subsec:approc_parent}, we examine the R2R score landscape and finite-search success using 100 simulation replicates for each $n\in\{250,500,1000\}$. 
Table~\ref{tab:r2r_search_diagnostics} summarizes the results, with all entries reported as percentages. Local-mode frequency is the proportion of replicates whose score landscape contains at least one non-global local mode. Global-basin coverage is the proportion of the $5{,}040$ starting orderings in basins of global optima. We also evaluate finite-search success for the stepwise objective using $2p^2=98$ proposals per start, reporting single-start and estimated best-of-five-start success rates. The stepwise approximation has little effect on local-mode frequency or global-basin coverage, while multiple initializations substantially increase the estimated probability of reaching a
global optimum.

\subsubsection{Comparison of ordering proposal operators}
\label{subsec:operators} 
We next examine how the choice of proposal operator affects hill climbing over variable orderings. Following the comparison considered by~\cite{chang2025identifiability}, we compare the R2R operator with two standard alternatives from the literature: random transposition (RTS) and adjacent transposition (ADJ). The RTS operator swaps two randomly selected positions, whereas ADJ swaps two neighboring positions. We consider simulated datasets with approximately $50\%$ zeros, $p=10$, $K=1$, and $n=1000$. For each proposal operator, we run the hill climbing algorithm using the same initialization and iteration budget. To evaluate search behavior, we plot the trajectory of the penalized score over iterations in Figure~\ref{fig:traj_plot}. It shows that R2R quickly converges to the score corresponding to the true ordering, whereas ADJ consistently becomes trapped in local modes. RTS achieves performance comparable to R2R. 

\begin{figure}[h!]
    \centering
    \captionsetup{font={normalsize,stretch=1}}
    \begin{subfigure}[b]{0.30\linewidth}
        \centering
        \includegraphics[width=\linewidth]{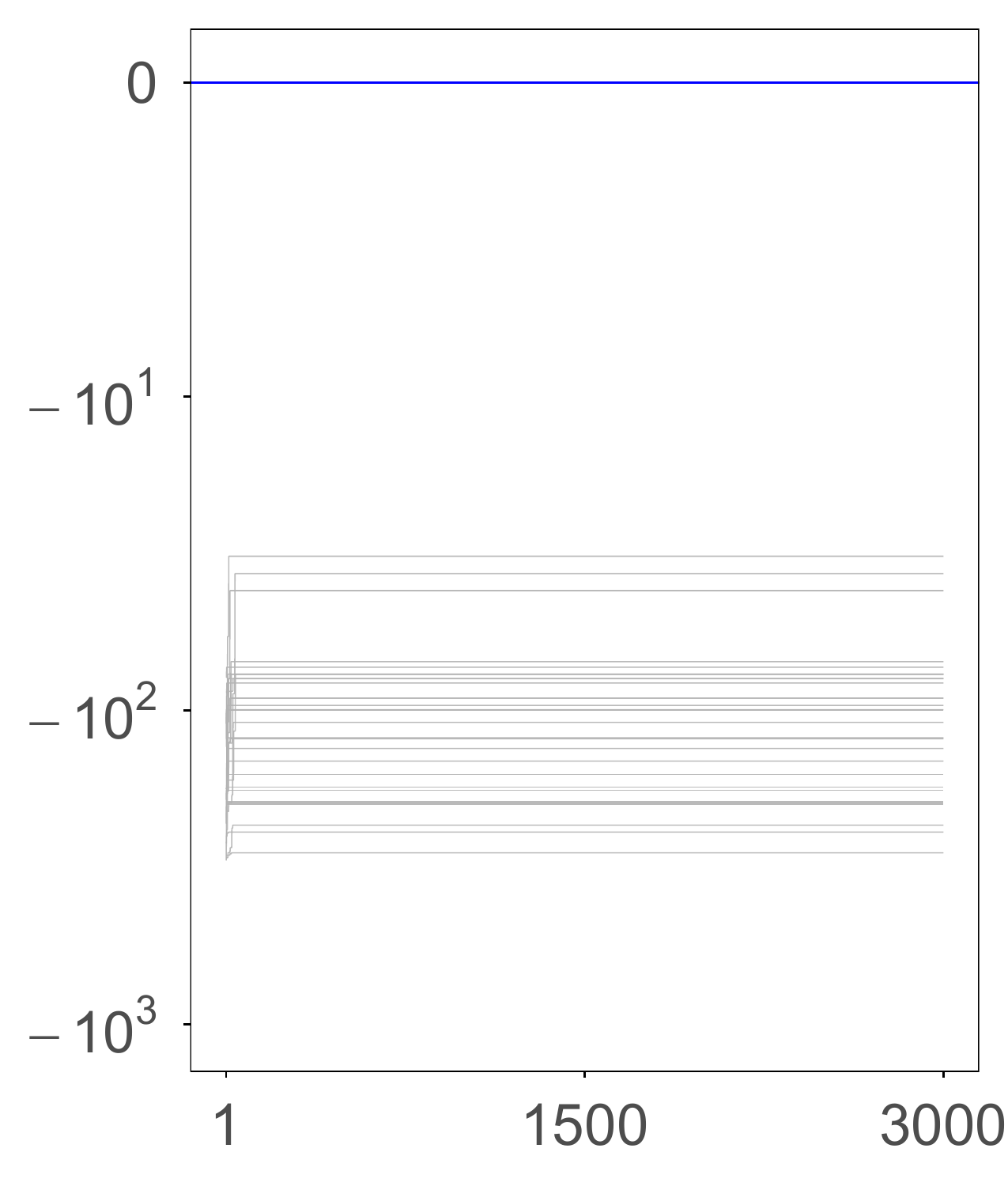}
        \caption{}
        \label{fig:neighbor_a}
    \end{subfigure}
    \hfill
    \begin{subfigure}[b]{0.30\linewidth}
        \centering
        \includegraphics[width=\linewidth]{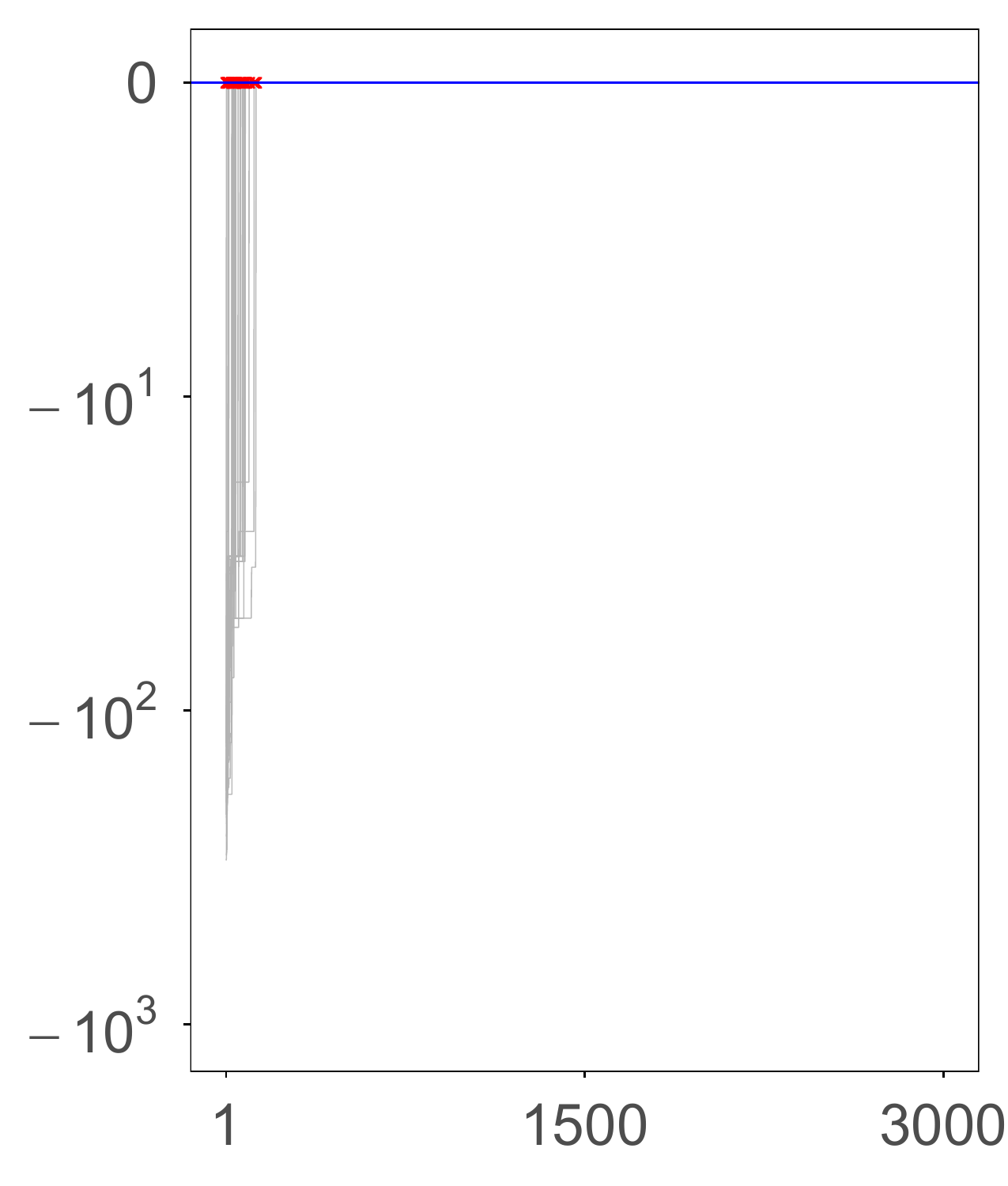}
        \caption{}
        \label{fig:neighbor_b}
    \end{subfigure}
    \hfill
    \begin{subfigure}[b]{0.30\linewidth}
        \centering
        \includegraphics[width=\linewidth]{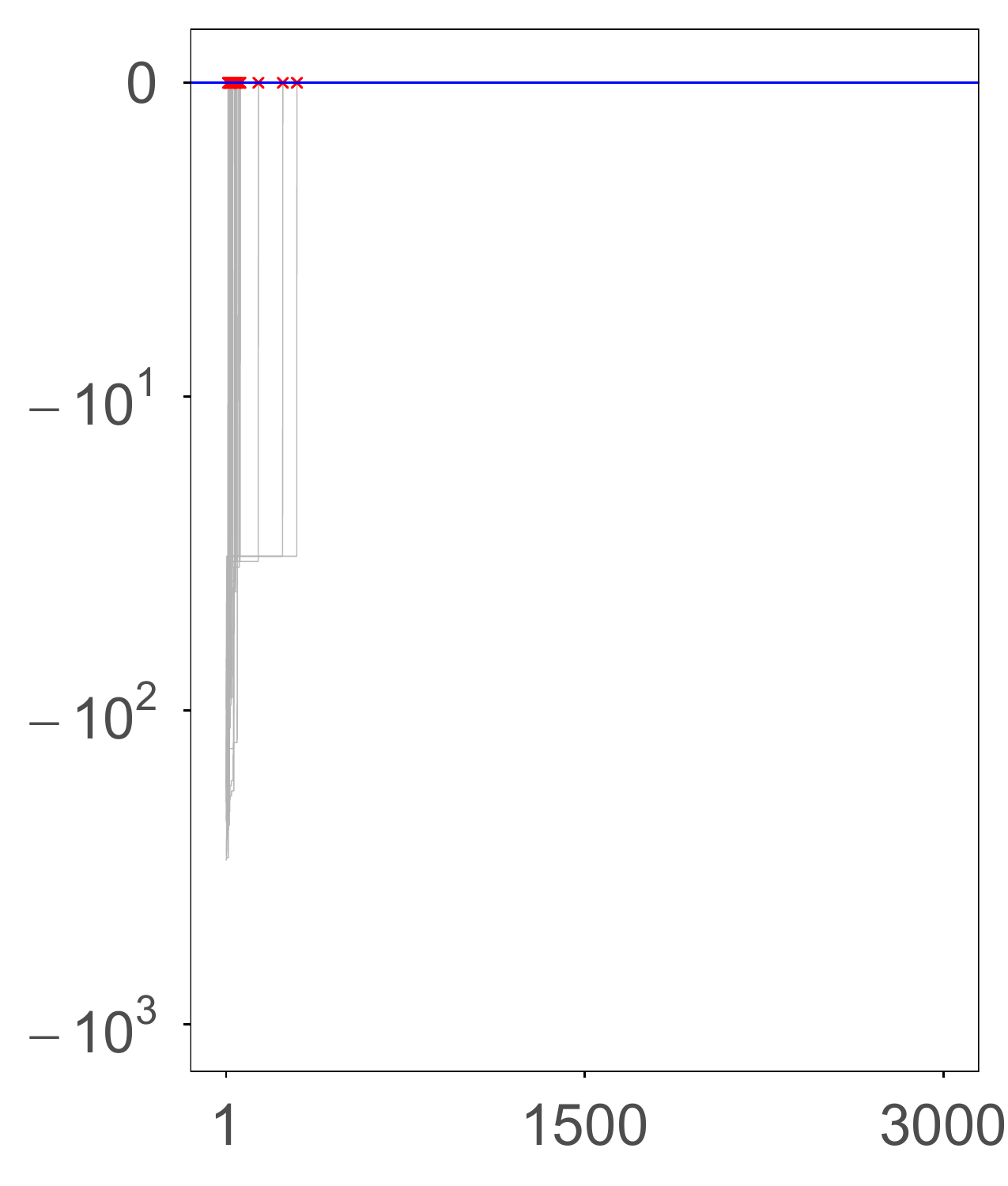}
        \caption{}
        \label{fig:neighbor_c}
    \end{subfigure}

    \caption{Scaled score versus the number of iterations for 30 runs with random initialization, with $p=10$ and $K=1$. The blue line represents the scaled score of the true ordering. Each red cross marks the first iteration at which the search reaches the reference score. Panels (a), (b), and (c) correspond to Algorithm~\ref{alg1} with the adjacent transposition (ADJ), random transposition (RTS), and random-to-random (R2R) operators, respectively.}
    \label{fig:traj_plot}
\end{figure}


\subsection{Robustness to modeling assumptions}
\label{subsec:assumption_sensitivity}

\subsubsection{Violations of the common ordering assumption}\label{subsec:common} 
\textbf{Interpretation.} The proposed joint estimation procedure assumes that related datasets share a common ordering structure. When the common ordering assumption is violated, simultaneous recovery of all true DAGs is not generally possible. In this case, we can regard the estimated ordering as a projection ordering, which maximizes the sum of the dataset-specific expected scores under the joint criterion.

\textbf{Sensitivity analysis.} To assess robustness to violations of this assumption, we generate datasets whose true orderings deviate from a common reference ordering. We use $p=10$, $K=5$, and $n=1000$ under the zero-inflated count data generating mechanism with approximately $50\%$ zeros, as described in Section~\ref{Section:Simulation 1}. Let $\sigma^*=(1,\ldots,p)$ denote a reference ordering. For each setting, we independently generate dataset-specific orderings $\{\sigma^{(k)}\}_{k=1}^K$ with a prescribed Kendall rank correlation relative to $\sigma^*$, and quantify their mutual agreement using the average pairwise Kendall rank correlation
\[
U =
\frac{2}{K(K-1)}
\sum_{1 \leq i < j \leq K}
\tau_{\mathrm{Kendall}}\!\left(\sigma^{(i)},\sigma^{(j)}\right).
\]
We consider agreement levels with average pairwise Kendall
correlations approximately equal to $U \in \{0.1,0.3,0.4,0.7,0.9,1.0\}$, where $U=1$ corresponds to the common ordering setting, and smaller values indicate stronger departures from a common ordering. For each value of $U$, we generate $K$ DAGs, where $G^{(k)}$ is consistent with the corresponding ordering $\sigma^{(k)}$, and simulate zero-inflated count data. The proposed method is then applied under the working assumption of a common ordering. We repeat each setting 30 times and summarize graph recovery using the average Hamming distance $\Delta$.
\begin{figure}[t!]
\centering
\captionsetup{font={normalsize,stretch=1}}
   \scalebox{0.90}{ 
    \begin{minipage}{0.55\textwidth}  
            \includegraphics[width=\linewidth]{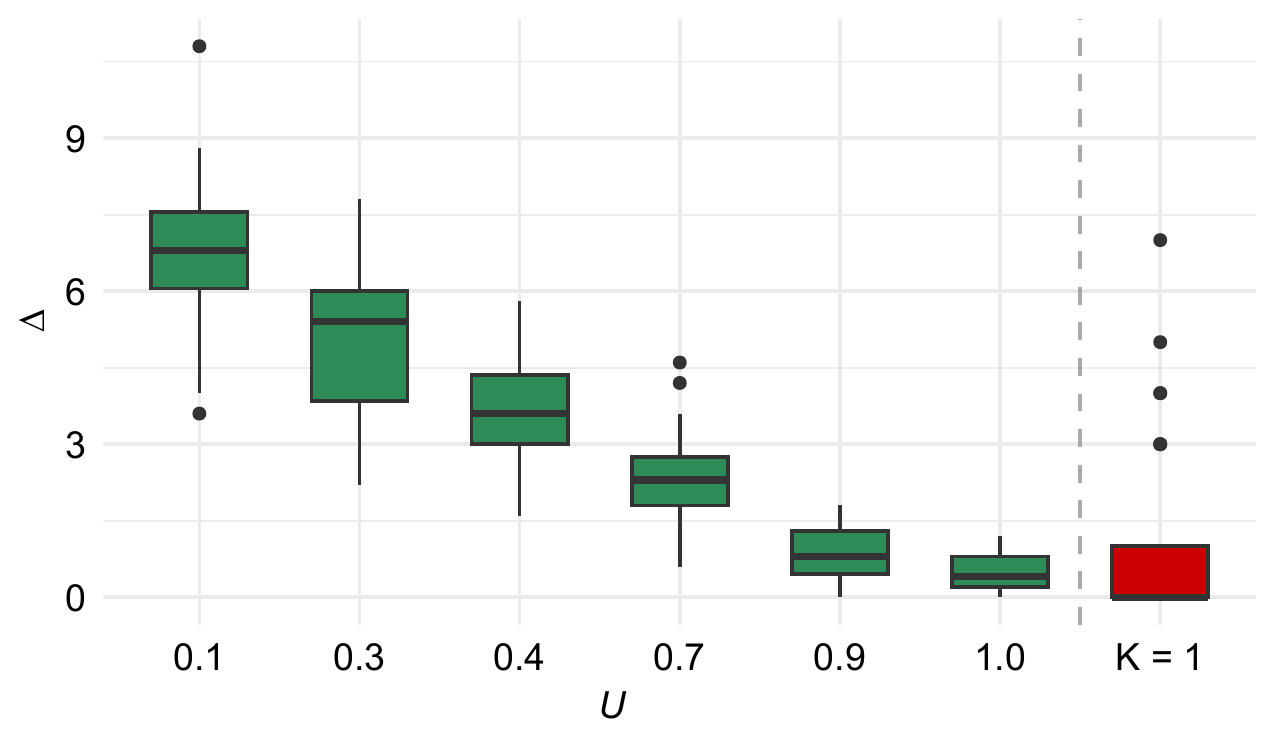}  
    \end{minipage}%
    }
    \caption{The green boxplots summarize the average difference in edges $\Delta$ between the true DAGs and their estimates using Algorithm~\ref{alg1} across different values of $U$, based on 30 experiments. The red boxplot corresponds to the case for $K = 1$.}
    \label{fig:common_order_violation}

\end{figure}
Figure~\ref{fig:common_order_violation} shows that recovery improves as the ordering agreement $U$ increases. When the dataset-specific orderings are highly aligned, especially for $U\ge 0.9$, joint estimation performs better on average than the individual-estimation benchmark. In contrast, when the common ordering assumption is substantially weakened, particularly for $U\le0.7$, individual estimation achieves a smaller Hamming distance. However, the individual-estimation benchmark also exhibits larger variability, whereas the joint method remains comparatively stable. These results suggest that the proposed method is robust to mild violations of the common ordering assumption, but substantial ordering disagreement can reduce the benefit of joint estimation.

\subsubsection{Misspecification of the conditional distribution} 
\label{subsec:model_misspecification}
Under distributional misspecification, Theorem~\ref{thm:consistency} no longer applies. We assess sensitivity to the ZIP conditional distribution by fitting ZIP and ZINB models to data generated from each family. To this end, we implement a ZINB version using the same order-search procedure and compare ZIP and ZINB fitted models with $n=1000$, $p=25$, and $K=1$. We follow the simulation setting of~\citet{choi2020bayesian}, generating data with approximately $50\%$ zeros and drawing the ZINB dispersion parameters independently from $\mathrm{Unif}(1,5)$.
Table~\ref{tab:model_misspecification} shows that fitting ZIP to ZINB-generated data substantially degrades graph recovery compared with fitting ZINB, whereas BIC favors the correctly specified likelihood under both generating distributions. 
\begin{table}[!h]
\centering
\captionsetup{font={normalsize,stretch=1}}
\caption{Performance of ZIP and ZINB fitted models under ZIP- and ZINB-generated data, with $n=1000$, $p=25$, and $K=1$. Entries report means (standard errors) over 30 simulation replicates.}
\label{tab:model_misspecification}
\small
\begingroup
\renewcommand{\arraystretch}{1.3}
\resizebox{\linewidth}{!}{%
\begin{tabular}{cccccc} 
\toprule
 Generating distribution & Fitted model & TPR & FDR & $\Delta$ & BIC\\
\midrule
\multirow{2}{*}{ZIP}
 & ZIP &  0.917 (0.016) & 0.104 (0.014) & 4.77 (0.710) & 69671.380 (655.940) \\
 & ZINB & 0.930 (0.016) & 0.083 (0.013) & 3.90 (0.660) & 69825.940 (654.990) \\
 \midrule
 \multirow{2}{*}{ZINB}
 & ZIP & 0.719 (0.026) & 0.359 (0.029) & 17.57 (1.610) & 90133.540 (416.310)  \\
 & ZINB & 0.880 (0.016) & 0.117 (0.016) & 5.93 (0.780) & 84549.990 (270.410)  \\
\bottomrule
\end{tabular}%
}
\endgroup
\end{table}


\subsection{Computational efficiency and acceleration strategy}
\label{subsec:computational_cost}

\subsubsection{Wall time analysis}
\label{subsec:walltime}
We use one Dell PowerEdge R6625 compute node with dual AMD EPYC 9334 CPUs, 768 GB RAM, local 480 GB SSD storage, 25GbE networking, and HDR InfiniBand connectivity up to 200 Gb/s. The wall times for the five replicates from the experiments reported in Table~\ref{tab:performance} are summarized in Table~\ref{tab:runtime}.
\begin{table}[!h]
\captionsetup{font={normalsize,stretch=1}}
\caption{Average wall time in minutes for varying $K$ and $p$, with $n = 1000$ and approximately $50\%$ zeros. Each entry reports the mean (standard error) based on five simulation replicates. The values in parentheses in the column headers indicate the number of iterations.}
\label{tab:runtime}
\centering
\small
\renewcommand{\arraystretch}{1.3}
\begin{tabular}{ccccc}
\toprule
 & \multicolumn{4}{c}{$p$} \\
\cmidrule(lr){2-5}
$K$ & 5 (250) & 10 (500) & 20 (1500) & 40 (3000) \\
\midrule
1 & 0.867 (0.030) & 3.617 (0.249) & 44.049 (7.046) & 98.640 (8.160) \\
5 & 1.298 (0.066) & 8.919 (0.160) & 86.401 (11.173) & 398.607 (24.739) \\
10 & 1.500 (0.056) & 10.823 (0.892) & 102.061 (12.154) & 619.797 (32.817) \\
20 & 1.946 (0.136) & 12.041 (0.462) & 161.456 (27.855) & 784.108 (84.835) \\
\bottomrule
\end{tabular}
\end{table}

\subsubsection{Acceleration of parent selection and full order-search}
\label{subsec:shortlist}

\textbf{Computational bottleneck.} Each proposed ordering requires \(Kp\) node-wise ZIP parent-selection problems. These problems can be evaluated independently across nodes and datasets. However, the ordering updates remain sequential. Consequently, parallelization can reduce the cost of evaluating an ordering without removing the repeated parent-selection work within each search iteration. Under ideal parallelization and comparable node-wise workloads, the computational cost is approximately
\[
    T(p)\times C_{\mathrm{VS}}(p,n),
\]
where \(T(p)\) is the number of R2R search iterations and \(C_{\mathrm{VS}}(p,n)\) is the cost of one ZIP parent-selection problem. Since \(T(p)\) is hard to reduce, we investigate how to reduce \(C_{\mathrm{VS}}\).

\textbf{Acceleration strategies.} We evaluate acceleration in two stages. A fixed-ordering experiment isolates its effect on parent selection, and a full-search experiment assesses its effect on the complete estimation procedure.

\textbf{(1) Parent selection under a fixed ordering.} We use \(p=200\) and \(n=1000\) to compare the current stepwise procedure with an in-degree constraint \(d_{\mathrm{max}}\), screening with \(m=40\)~\citep{fan2010sure}, and a conditional-score shortlist motivated by score-test screening~\citep{rao1948tests}. The shortlist scores all remaining variables but fully fits only the top \(b\) candidates, followed by refitting with the original unpenalized ZIP likelihood and evaluation using BIC. Table~\ref{tab:shortlist_fixed_order} reports the average number of full ZIP fits, wall time speed-up, and graph-recovery performance for each strategy. Restricting \(d_{\max}\) alone provides no computational gain in this experiment because it limits search depth but not the number of candidates evaluated at each step. In contrast, the conditional-score shortlist substantially reduces the number of full ZIP fits and reduces wall time while maintaining comparable graph-recovery performance under the fixed ordering.
\begin{table}[!h]
\captionsetup{font={normalsize,stretch=1}}
\caption{Computational cost and graph-recovery performance of variable-selection strategies under a fixed ordering, with $p=200$ and $n=1000$. Speed-ups are relative to the current stepwise method.}
\label{tab:shortlist_fixed_order}
\centering
\small
\begingroup
\renewcommand{\arraystretch}{1.3}
\begin{tabular}{lrrrr}
\toprule
Method & Average fits & Speed-up & TPR (\%) & FDR (\%) \\
\midrule
Stepwise (current)             & 246.6 & 1.00  & 96.5 & 15.4 \\
Bounded stepwise ($d_{\max}=5$) & 241.9 & 0.93  & 96.5 & 14.6 \\
Screening ($m=40$)             & 142.5 & 3.36  & 96.0 & 13.1 \\
Shortlist ($b=10$)             & 21.9  & 13.29 & 96.5 & 12.7 \\
\bottomrule
\end{tabular}
\endgroup
\end{table}

\newpage
\textbf{(2) Shortlisting within full order-search.} We next compare stepwise parent selection with shortlisting in the full R2R procedure using \(n=1000\), \(K=20\), \(p=40\), and \(T=3000\). The shortlist uses \(b=10\) and \(d_{\max}=5\). Table~\ref{tab:full_search_shortlist} reports means and standard errors across five simulation replicates, with speed-up calculated as the ratio of mean wall times.
\begin{table}[!h]
\captionsetup{font={normalsize,stretch=1}}
\caption{Computational cost and recovery performance of full R2R order-search with $n=1000$, $K=20$, $p=40$, and $T=3000$. The shortlist uses $b=10$ and $d_{\max}=5$. Entries report means (standard errors) across five simulation
replicates. Speed-up is the ratio of mean wall times relative to stepwise selection.}
\label{tab:full_search_shortlist}
\centering
\small
\begingroup
\renewcommand{\arraystretch}{1.3}
\setlength{\tabcolsep}{3pt}
\begin{tabular}{lcccccc}
\toprule
Parent selection & Wall time (min) & Speed-up
& $\Delta$ & $\tau$ & TPR & FDR \\
\midrule
Stepwise
& 784.108 (84.835) & 1.00
& 3.680 (0.200) & 0.960 (0.005)
& 0.947 (0.006) & 0.039 (0.002) \\
Shortlist
& 299.733 (38.770) & 2.62
& 4.020 (0.316) & 0.951 (0.002)
& 0.939 (0.007) & 0.040 (0.004) \\
\bottomrule
\end{tabular}
\endgroup
\end{table}

\newpage
\section{Additional real data analyses} \label{sec:real_supp}

\subsection{Data description and gene selection} \label{subsec:gene_description}
We analyze gene expression profiles from individuals diagnosed with major depressive disorder (MDD) and healthy controls. Table~\ref{tab:samplesize} summarizes the sample size for each group. 
\begin{table}[h!]
\centering
\captionsetup{font={normalsize,stretch=1}}
\caption{Sample sizes for 16 healthy control datasets and 17 MDD case datasets.}
\small
\renewcommand{\arraystretch}{1.3}
\begin{tabular}{cc|cc|cc|cc}
\toprule
\multicolumn{4}{c}{\textbf{Control}} & 
\multicolumn{4}{c}{\textbf{Case}} \\
\cmidrule(lr){1-4} \cmidrule(lr){5-8}
\textbf{ID} & \textbf{Sample size} & 
\textbf{ID} & \textbf{Sample size} & 
\textbf{ID} & \textbf{Sample size} & 
\textbf{ID} & \textbf{Sample size} \\
\midrule
1 & 1795 & 9  & 2630 & 1 & 3267 & 10 & 1688 \\
2 & 1363 & 10 & 3491 & 2 & 2423 & 11 & 3007 \\
3 & 2487 & 11 & 1953 & 3 & 1661 & 12 & 2196 \\
4 & 1234 & 12 & 2144 & 4 & 2275 & 13 & 2825 \\
5 & 2511 & 13 & 1061 & 5 & 2561 & 14 & 2690 \\
6 & 2433 & 14 & 1387 & 6 & 2511 & 15 & 2523 \\
7 & 3523 & 15 & 2461 & 7 & 3757 & 16 & 2077 \\
8 & 1239 & 16 & 3288 & 8 & 1758 & 17 & 3228 \\
 &      &    &      & 9 & 3227 &    &      \\
\bottomrule
\end{tabular}
\label{tab:samplesize}
\end{table}

Table~\ref{tab:mdd_gene_list} reports the 41 genes used in the real data analysis. These genes were reported by \citet{nagy2020single}, using an FDR threshold of  0.10.  
\begin{table}[h!]
\centering
\small
\captionsetup{font={normalsize,stretch=1}}
\caption{List of 41 genes used in the MDD case-control analysis.}
\label{tab:mdd_gene_list}
\renewcommand{\arraystretch}{1.15}
\begin{tabular}{llll}
\toprule
\multicolumn{4}{c}{Gene symbols} \\
\midrule
KAZN       & PINK1     & ARL8A     & ADORA1 \\
KIF26B     & AC133680.1& ACAA1     & CAMK2N2 \\
HIVEP1     & TAPBP     & GRM4      & PRKAR1B \\
DOCK4      & TMSB4X    & PRAF2     & TCEAL2 \\
TUBB4B     & CEND1     & TMEM132A  & FADS2 \\
NRXN2      & KLC2      & PPME1     & CAMK1D \\
CISD1      & FAM66C    & PEBP1     & TSC22D1 \\
HSP90AA1   & GRIN2A    & ALDOA     & ZBTB4 \\
TOM1L1     & MAP2K7    & RAB11B    & CCDC130 \\
ZFP36      & RPS11     & PTOV1     & SEZ6L \\
APP        &           &           &       \\
\bottomrule
\end{tabular}
\end{table}

\subsection{Additional network summaries}\label{subsec:voting_thres}
\textbf{Shared directed edges.} Table~\ref{tab:shared_edges} reports 15 directed edges selected across all 33 subjects. 
\begin{table}[!h]
\centering
\caption{Directed edges shared by all 33 subjects
(16 controls and 17 MDD).}
\label{tab:shared_edges}
\small
\setlength{\tabcolsep}{8pt}
\renewcommand{\arraystretch}{1.15}
\begin{tabular}{lll}
\toprule
Directed edge & Directed edge & Directed edge \\
\midrule
CAMK1D $\rightarrow$ DOCK4 &
GRIN2A $\rightarrow$ KAZN &
GRIN2A $\rightarrow$ CAMK1D \\

CAMK1D $\rightarrow$ TSC22D1 &
GRIN2A $\rightarrow$ PRKAR1B &
GRIN2A $\rightarrow$ TSC22D1 \\

CAMK1D $\rightarrow$ APP &
GRIN2A $\rightarrow$ DOCK4 &
GRIN2A $\rightarrow$ TOM1L1 \\

PEBP1 $\rightarrow$ TMSB4X &
GRIN2A $\rightarrow$ TCEAL2 &
GRIN2A $\rightarrow$ APP \\

TSC22D1 $\rightarrow$ APP &
GRIN2A $\rightarrow$ NRXN2 &
TOM1L1 $\rightarrow$ CAMK1D \\
\bottomrule
\end{tabular}
\end{table}

\textbf{Consensus networks.} We provide supplementary summaries of the group-specific consensus networks through edge counts across voting thresholds and out-degree rankings at \(\ell=8\). For group $g$ and voting threshold $\ell$, define the estimated edge set of the group-specific consensus network as
\[
\widehat E_g^{\mathrm{con}}(\ell)
=
\left\{
i\to j:
\sum_{k\in g}
\ind\!\left\{i\to j\in\widehat G^{(k)}\right\}
>\ell
\right\}.
\]
An edge is retained if it appears in more than $\ell$ subject-specific DAGs. Increasing $\ell$ gives a more stringent summary of cross-subject consistency.  The left panel of Figure~\ref{fig:edge_rank} reports the numbers of retained, shared, and group-specific edges across thresholds. The retained edge count in each group is non-increasing by construction.

\begin{figure}[!t]
    \centering
    \captionsetup{font={normalsize,stretch=1}}
    \includegraphics[width=0.49\linewidth]{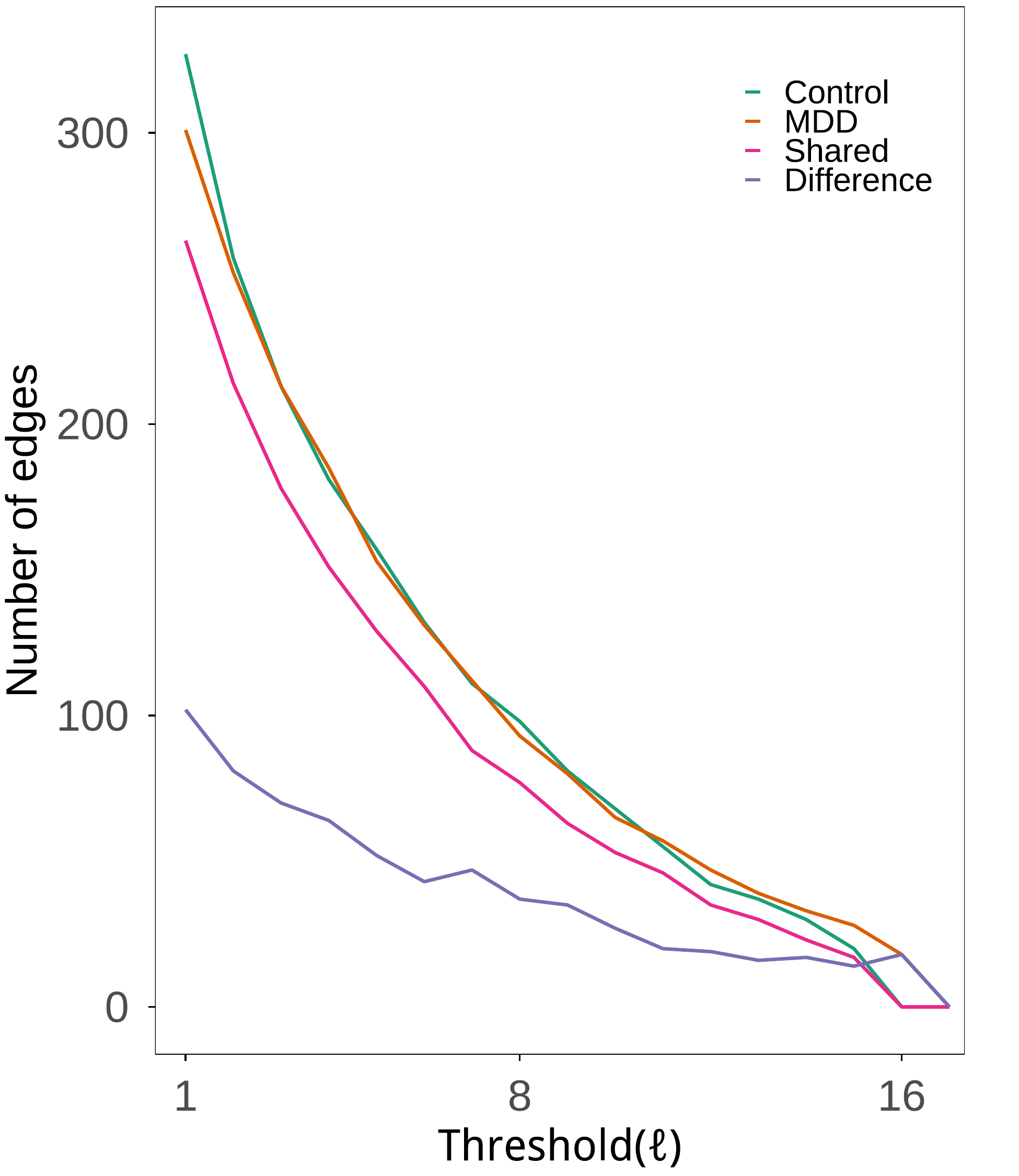}
    \includegraphics[width=0.49\linewidth]{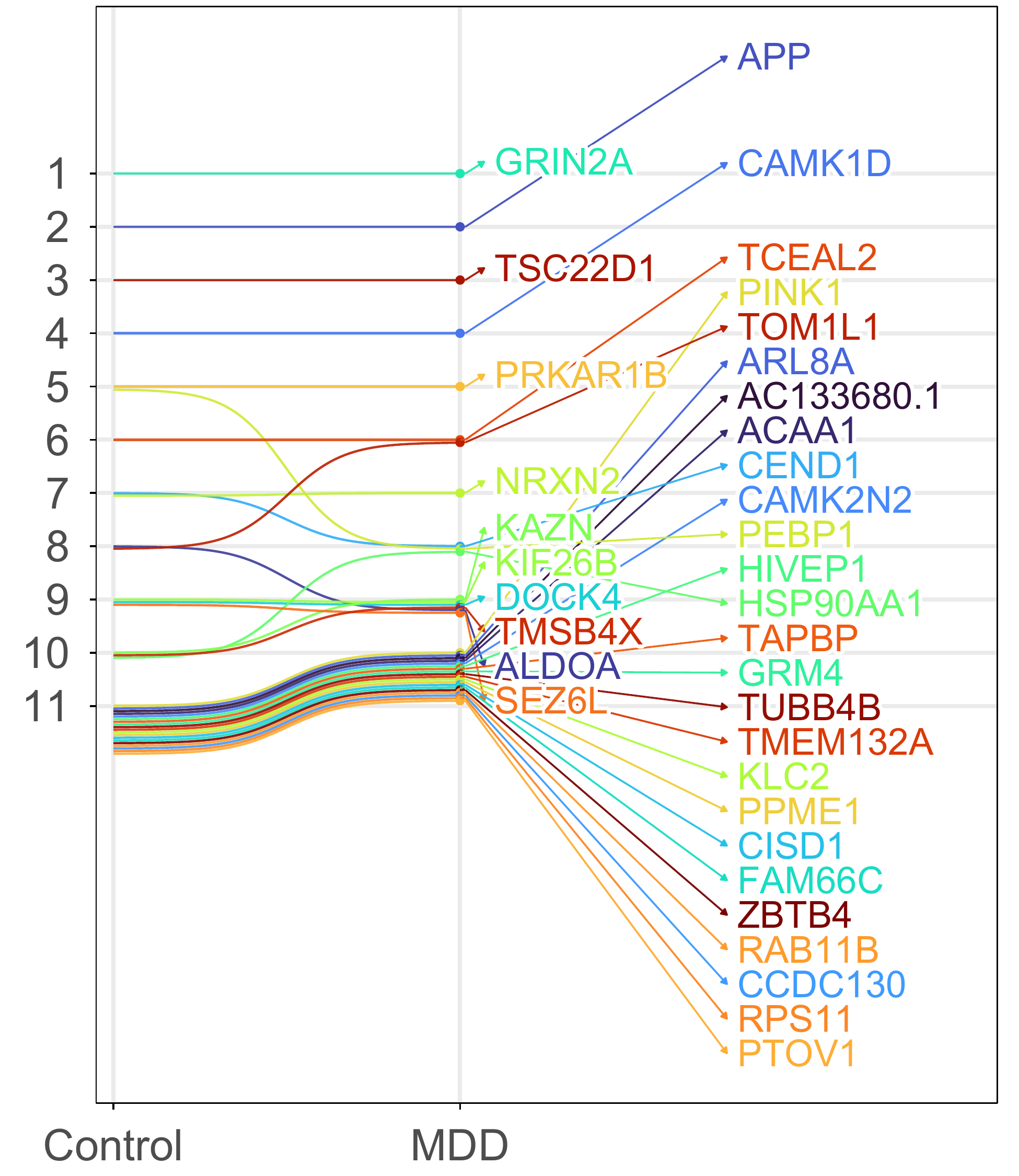}
    \caption{(Left) Number of edges in the consensus networks for controls, cases, their shared edges, and differences versus the threshold \(\ell\). (Right) Out-degree rank plots (with rank = 1 indicating the highest connectivity) illustrate how the estimated network connectivity shifts between the control and MDD groups.}
    \label{fig:edge_rank}
\end{figure}

\textbf{Out-degree ranking.} At \(\ell=8\), we rank genes separately within each group by decreasing out-degree, assigning the same rank to genes with equal out-degree. Rank 1 indicates the largest number of retained outgoing edges. The right panel of Figure~\ref{fig:edge_rank} presents these rankings as a complement to the total-degree rankings in Section~\ref{sec:real_data}.

\subsection{Compatibility with the common ordering} \label{subsec:shared_order}

The joint analysis assumes a common ordering across all subject-specific DAGs. To assess this assumption, we fit each subject separately and calculate the proportion of directed edges consistent with the ordering estimated from the joint analysis. The mean compatibility across subjects is 0.753, indicating substantial, though not complete, directional agreement. Disagreement may reflect finite-sample variability, unmeasured regulators, or feedback mechanisms not captured by a DAG. As we illustrate in Section~\ref{subsec:common}, we may interpret the result as a useful approximation rather than an actual biological constraint. 

\end{document}